\documentclass[aps,pre,twocolumn,groupedaddress]{revtex4-2}
\usepackage{graphicx}
\usepackage{amsmath,amssymb,bm}

\usepackage{CJK}
\begin{document}
\begin{CJK*}{UTF8}{ipxm}

\title{X-ray computerized tomography observation of Lycopodium paste incorporating memory of shaking}


\author{So Kitsunezaki (狐崎 創)}
\homepage[]{http://www.noneq.phys.nara-wu.ac.jp/~kitsune}
\email[]{kitsune@ki-rin.phys.nara-wu.ac.jp}
\affiliation{Research Group of Physics, Division of Natural Sciences, 
 Faculty of Nara Women's University, Nara 630-8506,Japan}

\author{Akihiro Nishimoto (西本 明弘)}
\affiliation{Faculty of Health and Well-being, Kansai University, Sakai 590-8515, Japan}

\author{Tsuyoshi Mizuguchi (水口 毅)}
\affiliation{Department of Physics, Osaka Prefecture University, Sakai 599-8531, Japan}

\author{Yousuke Matsuo (松尾 洋介)}
\author{Akio Nakahara (中原 明生)}
\affiliation{Laboratory of Physics, College of Science and Technology, Nihon University, Funabashi 274-8501, Japan}



\date{\today}

\begin{abstract}
In a uniform layer consisting of a mixture of granular material and liquid, it is known that desiccation cracks exhibit various anisotropic patterns that depend on the nature of the shaking that the layer experienced before drying. 
The existence of this effect implies that information regarding the direction of shaking is retained as a kind of memory in the arrangements of granular particles. 
In this work we make measurements in paste composed of Lycopodium powder using microfocus x-ray computerized tomography ($\mu$CT) in order to investigate the three-dimensional arrangements of particles. 
We find shaking-induced anisotropic arrangements of neighboring particles and density fluctuations forming interstices mainly in the lower part of the layer. 
We compare the observed properties of these arrangements with numerical results obtained in the study of a model of non-Brownian particles under shear deformation. 
In the experimental system, we also observe crack tips in the $\mu$CT images and confirm that these cracks grow along interstices in the direction perpendicular to the initial shaking.   
\end{abstract}


\maketitle
\end{CJK*}

\section{Introduction\label{sec:introduction}}

Mixtures of fine solid particles and liquid (for example, water) behave as viscous fluids with plasticity. 
Such mixtures are generally referred to as granular pastes. 
It has been found that cracks often appear in granular pastes as they transition to semisolid states during desiccation \cite{Coussot05,Goehring15}.  
When a uniform layer of granular paste dries evenly on a hard flat surface, cracks usually form an isotropic cellular pattern. However, previous studies have revealed that the crack patterns can be made anisotropic if a mechanical or electromagnetic perturbation is applied to the layer for a short time before drying. 
This memory effect of paste suggests that we can readily impart anisotropic mechanical properties on pastelike materials and control the process of cracking \cite{Nakahara05,Nakahara06,Nakahara06b,Nakahara11,Matsuo12,Nakayama13,Goehring15}.  
However, our understanding of such behavior is yet incomplete, mainly because we do not know how external perturbations generate structures incorporated as memory in paste and how these structures influence cracking processes.   

The memory of shaking is the most common type of memory effect, being observed in many types of paste.
In situations that paste exhibits this effect, after horizontal unidirectional oscillation is applied to a uniform layer of paste, parallel cracks first appear during desiccation in the direction perpendicular to the initial shaking. 
This effect was first found in claylike paste consisting of calcium carbonate \cite{Nakahara05} and then also observed in wet granular materials, such as mixtures of starch powder and water \cite{Kitsunezaki17}. 
As these experimental studies revealed, parallel cracks form only when the shaking is sufficiently strong to exert shear stresses larger than the yield stress on the paste. 
Further, it was found that many types of crack patterns, including rings and spirals, can be produced by changing the manner of shaking \cite{Nakahara05,Nakahara06b,Nakayama13}.
Recent experiments have also indicated that such memory can be rewritten by adding only a few oscillations in a direction that differs from that of the initial shaking \cite{Nakahra19}. 
Nonlinear elastoplastic theories provide predictions that are consistent with these results. 
The phenomenological models proposed by Otsuki and Ooshida predict that large shear deformation yields anisotropic residual stresses, and this prediction has been confirmed by measurements of stresses in calcium carbonate paste \cite{Otsuki05,Ooshida08,Ooshida09,Kitsunezaki16,Morita21}. 

What anisotropic structures are formed microscopically by shaking and how do they influence the fracture properties of paste? 
In a previous study \cite{Kitsunezaki17} we observed the arrangements of particles in several samples using x-ray computerized tomography ($\mu$CT) and found that shaking induces short-range anisotropy in which the number of neighboring particles increases in the direction perpendicular to the direction of shaking in the shear plane. 
However, we were not able to investigate the details of the anisotropy and their relation to the resultant cracks. 
In this work we use a $\mu$CT apparatus at the large synchrotron radiation facility SPring-8 (RIKEN, Japan). 
With this apparatus, we are able to realize a field of view with a diameter of $3.56 \mathrm{mm}$, which is approximately four times larger than that used in the previous study, while maintaining the same resolution. We carry out measurements in a few dozen samples. 
As in the previous study, we use paste consisting of Lycopodium powder, i.e., spores of {\it Lycopodium clavatum}. 
These particles have round shapes with diameters of approximately $30\;\mathrm{\mu m}$. 
These particles are the largest among the spherical particles whose paste is known to exhibit memory of shaking, and due to this large size, individual particles can be resolved in the $\mu$CT observations. 

In this study, we confirm that such short-range anisotropy is induced only when the sample is shaken under the conditions that the memory effect appears and that it is pronounced in the lower part of the layer. 
We also find that shaking induces not only short-range anisotropy but also anisotropic fluctuations in the density of particles forming interstices.   
We find that this anisotropic density fluctuation is a direct cause of the memory effect.  
By visualizing the vicinities of growing crack tips, we determine its relation to the crack growth direction.

In the next section we explain the methods of sample preparation and the $\mu$CT measurements. In Sec.~\ref{sec:orderparams} we analyze the height dependence of the directional order parameters of the arrangement of particles. 
In Sec.~\ref{sec:interstices} we present results for the properties of interstices and their relation to the direction of crack growth. 
In Sec.~\ref{sec:discussion} we discuss numerical simulations of non-Brownian particles under shear deformation. 
In these simulations, we find anisotropy in the arrangement of particles similar to that seen in the experiments. 
In Sec.~\ref{sec:conclusion} we give a summary.

\section{Methods\label{sec:methods}}

We prepared a layer consisting of a mixture of Lycopodium powder and cesium chloride (CsCl) solution containing a small amount of agar and applied horizontal shaking.  
After drying the paste in a high-temperature environment of $45.0\mathrm{{}^\circ C}\pm 1.5\mathrm{{}^\circ C}$ for a predetermined period, we covered it to stop desiccation and then allowed it to solidify through gelation of the agar at room temperature. 
We prepared numerous samples under various conditions in advance and  cut out pieces of a layer from each sample for $\mu$CT observation at SPring-8. 
The main parameters characterizing our experiments are the frequency of shaking $f$, the initial solid volume fraction, $\phi(0)$, and the solid volume fraction at the time of gelation $\phi_g$. 

We prepared every sample such that the area density of Lycopodium powder (Association of Powder Process Industry and Engineering, Japan) was $0.070\;\mathrm{g/cm^2}$ and we controlled the initial solid volume fraction to take values between $\phi(0)=10\%$ and $23\%$ by adjusting the volume of the solution. 
The solution contained $140\;\mathrm{g}$ of CsCl (Wako Pure Chemical Industries, Osaka, Japan) and $6.00\;\mathrm{g}$ of agar (Ina Food Industry, Japan) per liter of ion-exchanged water as radiopaque and gelation agents, respectively. (The solidifying temperature of the agar was approximately equal to $40\mathrm{^{\circ}C}$.
\footnote{The density of the agar was increased from the value $4\;\mathrm{g/l}$ used in Ref.~\cite{Kitsunezaki17} in order to solidify samples with smaller solid volume fractions in the initial stages of desiccation.})
The solid volume fraction of the paste was determined from the densities of the CsCl solution ($1.10\;\mathrm{g/cm^3}$) and a Lycopodium particle ($1.05\;\mathrm{g/cm^3}$) by assuming the small amount of agar to be negligible.
Lycopodium powder was mixed with the solution using a hot magnetic stirrer and then stored in a sealed flask at $70\;\mathrm{{}^\circ C}$ until it was poured into containers.

\begin{figure*}[t]
\includegraphics[width=16.0cm]{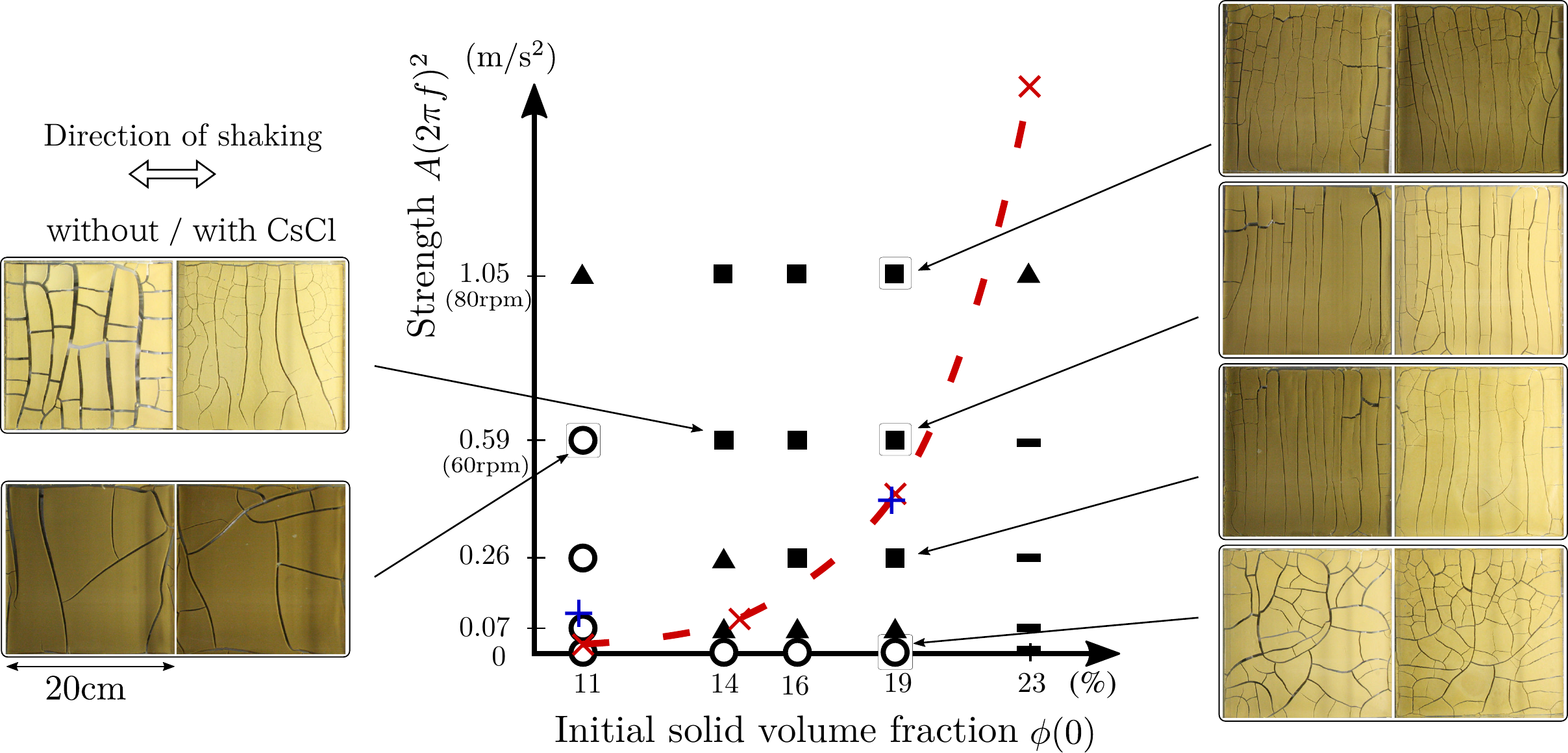} 
\caption{
Morphological diagram of desiccation cracks in paste consisting of Lycopodium powder and $6\;\mathrm{g/l}$ of agar solution. 
The values on the vertical axis correspond to $f=0, 20, 40, 60$, and $80\;\mathrm{rpm}$ from bottom to top. 
The photographs display typical crack patterns found in a paste sample contained in a $20\times 20\;\mathrm{cm^2}$ acrylic container with the various sets of conditions indicated by the arrows. 
The photographs on the left- and right-hand sides are of systems without CsCl and with a $140\;\mathrm{g/l}$ density of CsCl, respectively.
The open circles indicate the appearance of isotropic crack patterns, the squares indicate the formation of cracks in lines perpendicular to the direction of shaking, and the triangles indicate the appearance of partially anisotropic patterns. 
Under the conditions indicated by the minus signs, paste did not spread during shaking, due to large yield stresses. 
For this reason, we could not prepare a uniform layer for experiments. 
The open squares indicate the conditions on which the present work is focused. 
The dashed curve, the red crosses, and the blue pluses represent the results of the measurements of yield stresses discussed in the text. 
\label{fig:phasediagram}}
\end{figure*}

The existence of the memory effect in a paste of Lycopodium powder and agar solution was confirmed by pouring paste into containers and then drying it in the high-temperature environment until crack formation, as shown in Fig.~\ref{fig:phasediagram}. 
We found that the morphological diagram of crack patterns obtained in this case is qualitatively the same as in the case without agar, although the plastic limit decreases by approximately 3\% \cite{Kitsunezaki17}. 
We also found that adding CsCl to the agar solution does not change the crack patterns significantly, as seen in the photographs in Fig.~\ref{fig:phasediagram}
\footnote{However, we found that if we stored paste containing both CsCl and agar  for more than a few days before the experiments,  the memory effect of shaking was lessened and in some cases disappeared. 
In this case, the samples tended to become fragile and exhibit irregular crack patterns.  
For this reason, we used only fresh paste stored for less than $12\mathrm{h}$ in preparing samples.}.
In order to clarify the differences among the conditions, we mainly report the results obtained under the conditions corresponding to the open squares in Fig.~\ref{fig:phasediagram}. 

The yield stress curve of Lycopodium paste is plotted in Fig.~\ref{fig:phasediagram}. 
The yield stresses $\sigma_Y$ were measured by using a rheometer (Physica MCR301, Anton Paar, Austria).
These measurements were made on paste maintained at $45\;\mathrm{{}^\circ C}$ inside a coaxial double cylindrical vessel to which a stress with constantly increasing magnitude was applied. 
The value of $\sigma_Y$ was determined at the commencement of the first rotation using the least-squares method. 
For samples without CsCl, the measured values of $\sigma_Y$ were $0.14$, $0.43$, $1.5$ and $4.6\;\mathrm{Pa}$ for $\phi(0)=10.6\%, 14.5\%, 19.2\%$ and $22.9\%$, respectively,  and $\sigma_Y>200\;\mathrm{Pa}$ for $\phi(0)=25.2\%$, although we found that the values decreased by amounts in the range $0\%$--$30\%$ at the second rotation.

The crosses in Fig.~\ref{fig:phasediagram} indicate the strengths of shaking that generate the yield stresses on the bottom of a layer and the dotted curve indicates the yield stress curve determined from these data.  
Anisotropic crack patterns were observed clearly above this curve and near the plastic limit, as expected.  
The yield stresses of samples containing CsCl were found to be $0.8$ and $1.6\;\mathrm{Pa}$ for $\phi(0)=10.5\%$ and $19.1\%$, respectively. 
The corresponding shaking strengths are indicated by the pluses.  
Adding CsCl to the agar solution did not significantly affect the rheological properties of the paste, although the reproducibility of $\sigma_Y$ deteriorated for small yield stresses.

We applied a horizontal oscillation of amplitude $A=15\;\mathrm{mm}$ to a layer of paste using a shaker (FNX-220, TGK, Tokyo, Japan). 
We refer to the shaking direction as the $x$ axis and to the upward vertical direction during shaking as the $z$ direction in this paper.  
The applied frequency $f$ was varied from $0$ to $80\;\mathrm{rpm}$; in the case $f=0$, the paste experienced no shaking after being poured. 
Shaking was applied for $5\;\mathrm{min}$ immediately after the paste was poured into a container positioned on the shaker in the high temperature environment. 
The surface temperature of a layer of paste was in the range $50\;\mathrm{{}^\circ C}$--$60\;\mathrm{{}^\circ C}$ immediately after being poured and then it decreased to approximately the ambient temperature within $30\;\mathrm{s}$. 
Although we omit the results, in some samples that were covered after being poured, we did not observe the anisotropic structures reported in this paper.  
We believe that this resulted from the fact that, due to the insulating effect of the cover, these samples did not cool sufficiently and this resulted in different rheological properties during shaking. 

\begin{figure*}[t]
\includegraphics[width=16.0cm]{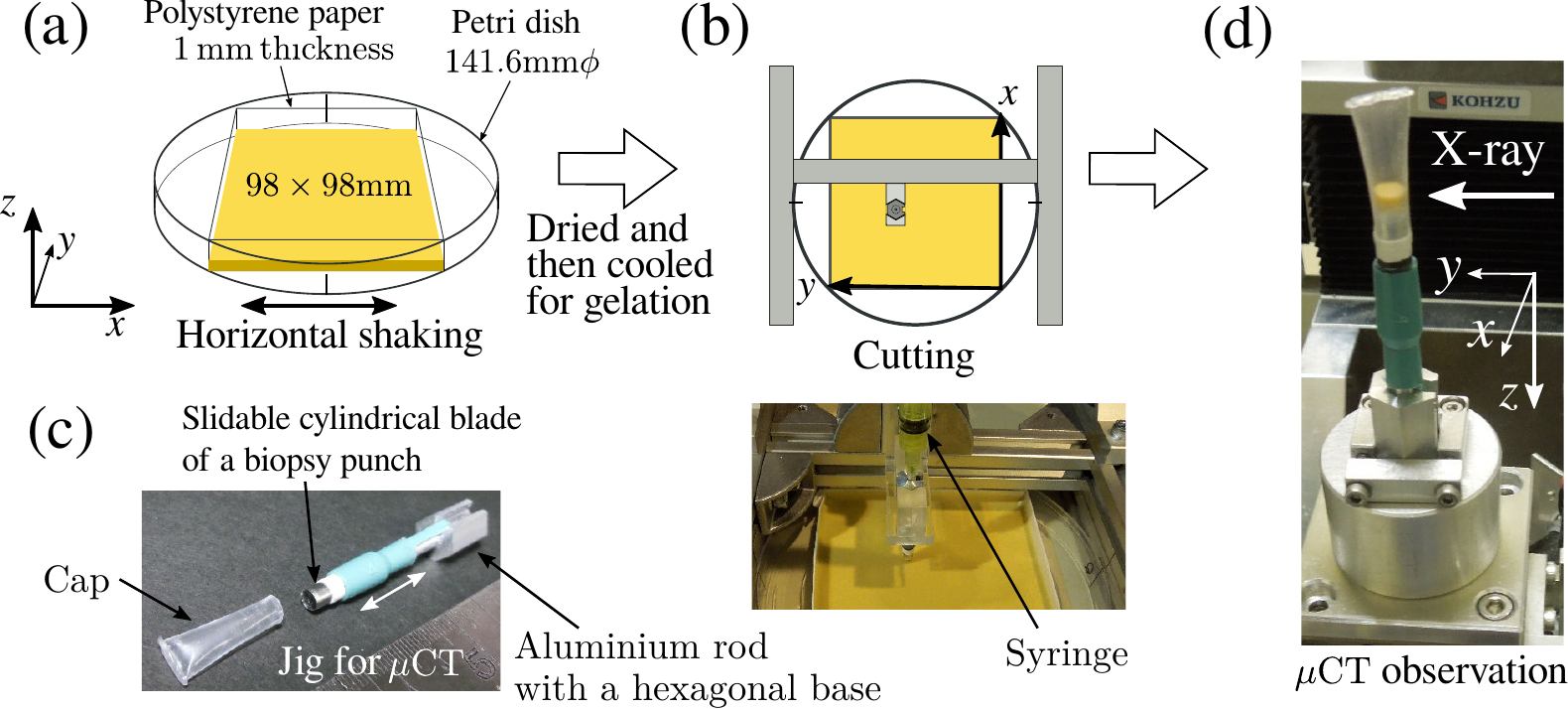} 
\caption{
Preparation of a sample for $\mu$CT observation. 
(a) Paste of Lycopodium powder was poured into a square container composed of thin polystyrene paper formed in a Petri dish. 
After we applied unidirectional horizontal shaking in a high-temperature environment, the paste was dried in the same environment and then allowed to solidify at room temperature. 
(b) We attached the jig displayed in (c) to an aluminum frame and cut out the layer of paste vertically with a cylindrical blade. 
(d) The jig was fixed upside down to the turntable for the $\mu$CT scanning.
\label{fig:preparation}
}
\end{figure*}

We prepared a sample for the $\mu$CT measurements by pouring paste into a square container of side length $98\;\mathrm{mm}$, which was constructed from polystyrene paper attached to a circular plastic Petri dish, as depicted in Fig.~\ref{fig:preparation}(a). 
The use of polystyrene paper made it easy to remove a cut piece of a layer from the bottom of a container.  
We prepared many solidified samples under various conditions, i.e., various values of $(f,\phi(0),\phi_{g})$, during a period from approximately one week to one day before the $\mu$CT measurements.  
The samples were then transported to SPring-8 and measurements were carried out during a $24$-$\mathrm{h}$ period assigned in advance by SPring-8. 
We performed such measurements three times several months apart.

\begin{figure*}[t]
\includegraphics[width=17.5cm]{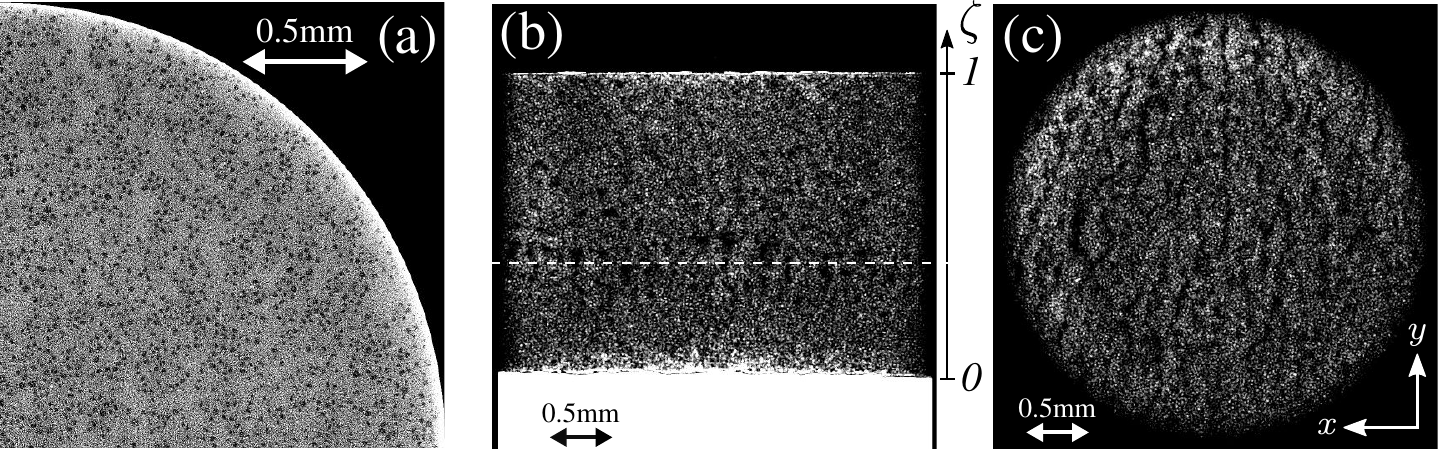} 
\caption{
Typical $\mu$CT images of a paste with memory of shaking. 
This sample was prepared with $(f,\phi(0),\phi_g)=(60\;\mathrm{rpm}, 18.4\%, 24.0\%)$. 
The arrows indicate both the direction of shaking and the length scale corresponding  to $0.5\;\mathrm{mm}$. 
(a) A quarter of a horizontal cross section depicted in the original 3D grayscale. 
In this image, the CsCl solution appears brighter than the Lycopodium particles because of its large opaqueness with respect to x rays. 
(b) and (c) Vertical and horizontal cross sections after binarization. 
Here the particles are identified as white regions (the brightness outside the sample is not inverted). 
Running averages over $80$ cross sections ($80l_p=139.2\;\mathrm{\mu m}$) were performed in the depth direction of each image in order to make the particle arrangements clearly visible. 
The cross sections in (a) and (c) are displayed at the height indicated by the dotted line in (b). 
The height from the bottom of a layer is represented by $\zeta$, the fraction of the cumulative number of detected particles located below the given height. 
}
\label{fig:cross_section}

\

\includegraphics[width=11.5cm]{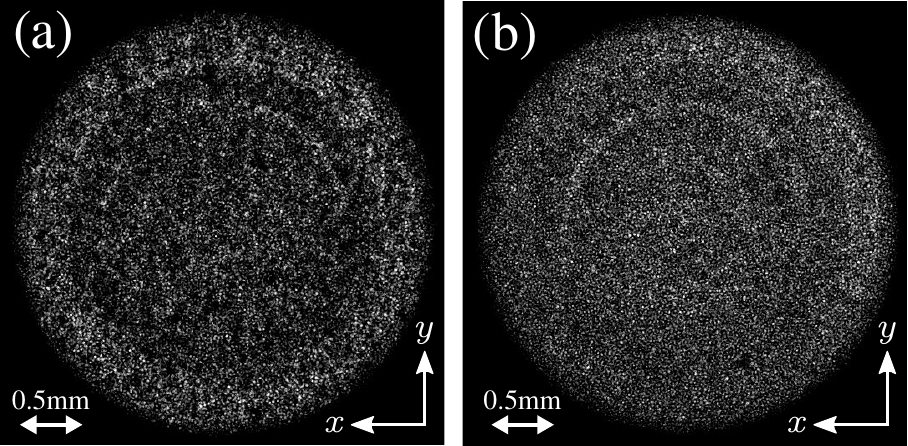} 
\caption{
Typical $\mu$CT images  of pastes without memory of shaking.  (a) The sample was prepared with a small initial solid volume fraction and shaken at the same frequency as that in Fig.~\ref{fig:cross_section}: $(f,\phi(0),\phi_g)=(60\;\mathrm{rpm}, 10.4\%, 18.6\%)$. (b) Unshaken case with the same initial solid volume fraction as that in Fig.~\ref{fig:cross_section}: $(f,\phi(0),\phi_g)=(0\;\mathrm{rpm}, 18.4\%, 25.3\%)$.  
These horizontal cross sections were taken using the same $\zeta$ and the same method as in Fig.~\ref{fig:cross_section}(c).
The ringlike unevenness in brightness is an artifact of the image construction.
}
\label{fig:cross_section_nomemory}
\end{figure*}

Our measurements were performed using x-rays of $25\;\mathrm{keV}$ from the BL20B2 beam line of SPring-8. 
One pixel in the $\mu$CT images corresponds to a region of length $l_p\equiv 1.74\;\mathrm{\mu m}$ in a sample. 
The field of view is a cylindrical region with both a diameter and height of $2048 l_p=3.56\;\mathrm{mm}$.  
For the measurements, we cut out a piece of a layer from each sample, avoiding the vicinities of cracks and the boundary of the container, as depicted in Fig.~\ref{fig:preparation}(b), except in the cases in which we observed the crack tips, considered in Sec.~\ref{sec:interstices}. 
In order to minimize the disturbance to the samples caused by the cutting process, we constructed a jig from an aluminum rod to which a cylindrical blade (diameter $4\;\mathrm{mm}$) of a biopsy punch (BPP-40F, Kai industries, Gifu, Japan) was attached in such a manner that it could be slid along a single direction, as shown in Fig.~\ref{fig:preparation}(c). 
The jig had a hexagonal base, which allowed us to fix the direction of each cut. 
We cut a sample vertically with the blade and gently removed the cut layer with the blade just prior to the $\mu$CT observation. 
Because very soft samples often became stuck, despite our use of the polystyrene paper, the jig was prepared with a narrow hole running its entire length, and through this hole we applied negative air pressure with a syringe to gently remove a sample. 
We fixed the jig upside down on the turntable for $\mu$CT scanning, and then, attaching a cap to prevent desiccation, we slid the blade down to place the cut layer on the top of the rod for x-ray irradiation. 
Each $\mu$CT measurement took $20$--$30\;\mathrm{min}$. 

We constructed three-dimensional (3D) images from the measurement data using programs provided by SPring-8 \cite{Uesugi}.  
Figure~\ref{fig:cross_section}(a) displays a typical image of a cross section. 
In the image, the brightness increases with the x-ray absorption rate and thus, in order of increasing brightness, we have regions consisting of air, Lycopodium powder, and gelated CsCl solution. 
After applying brightness inversion and noise reduction, we obtained images of particles through binarization \cite{Fiji}. 
Figures~\ref{fig:cross_section}(b) and \ref{fig:cross_section}(c) display typical vertical and horizontal cross sections of a sample prepared under typical conditions for which the paste exhibits memory of shaking.  

In these figures, it is seen that the density of particles is not uniform and low-density regions form interstices along the direction perpendicular to the applied shaking. 
We do not find such an anisotropic structure in samples without memory of shaking;  
Fig.~\ref{fig:cross_section_nomemory}(a) displays a typical horizontal cross section of a sample prepared with an initial volume fraction too small for memory of shaking to appear, while Fig.~\ref{fig:cross_section_nomemory}(b) corresponds to the unshaken case of Fig.~\ref{fig:cross_section}.  

We investigate the short-range anisotropy in the arrangement of neighboring particles in Sec.~\ref{sec:orderparams} and then features of these anisotropic interstices in Sec.~\ref{sec:interstices}.
We calculated the center of mass of the particles from the 3D binary images \cite{BoneJ}. 
As a measure of the height within a sample, we introduce the quantity $\zeta(z)$, representing the fraction of the total number of particles contained in the 3D image that exist below $z$, the height above the bottom.  
Because we took the 3D images such that the bottom of a layer was included and the concentration of particles was nearly constant throughout a sample, $\zeta(z)$ is approximately proportional to $z$. 
Generally, the position corresponding to $\zeta=1$ coincided with the top surface of a layer.
However,  there were a few samples with small $\phi_g$ (smaller than approximately $15\%$) for which  the entire height of the layer was not included in the 3D image due to the large thickness. 
In such cases, therefore, $\zeta = 1$ did not correspond to the top surface. 
The total number of Lycopodium particles detected in a 3D image was approximately $5\times 10^5$ when the entire height of the layer was included.

\section{Height dependence of directional order parameters\label{sec:orderparams}}

\begin{figure*}[t]
\includegraphics[width=14.0cm]{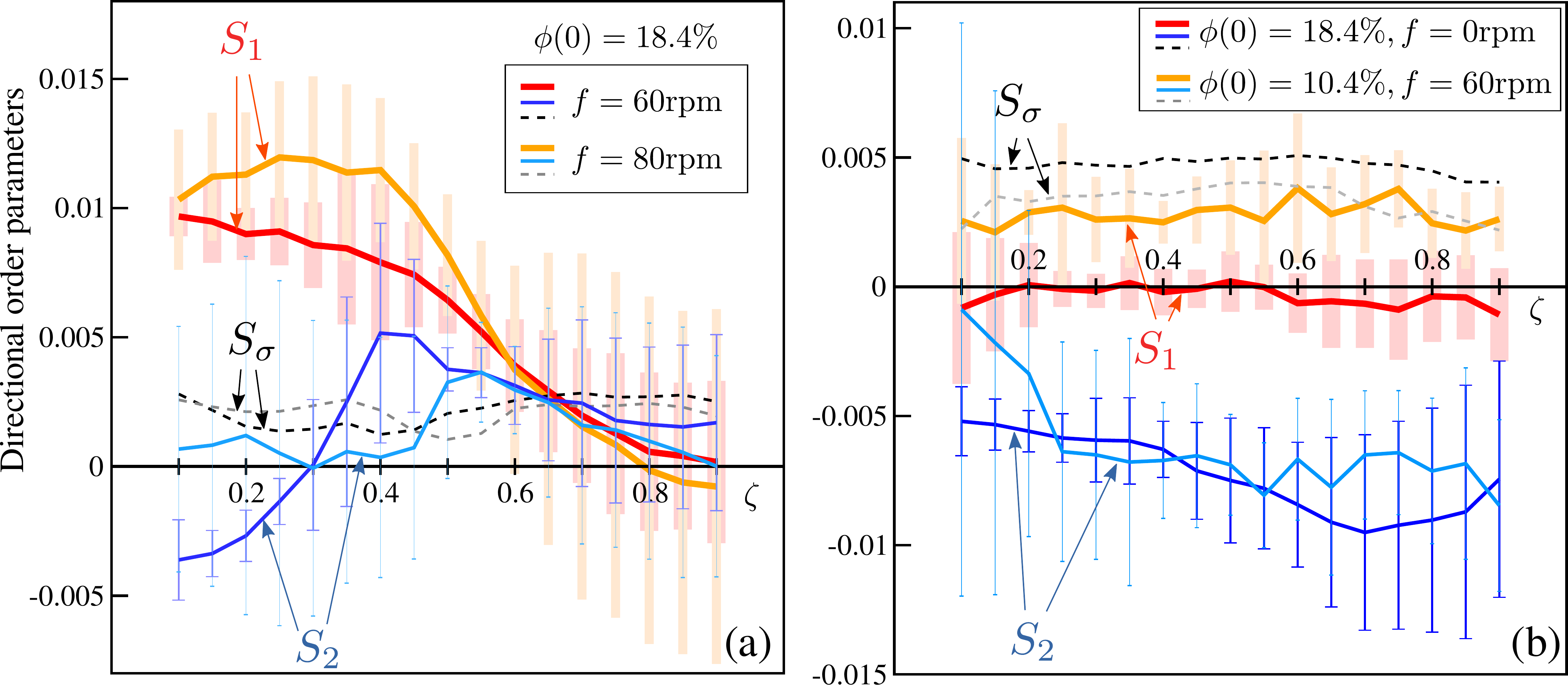} 
\caption{
Height dependences of the directional order parameters $S_1$ and $S_2$. 
The averages of $S_1$ and $S_2$ over five samples are plotted with the thick and thin solid curves, respectively, along with error bars, for conditions under which the paste (a) exhibits and (b) does not exhibit the memory effect of shaking.  
The quantity $S_{\sigma}$ [defined in Eq.~(\ref{S_1,S_2,S_sigma})] is indicated by the dotted curves.  
In each figure, the horizontal axis represents $\zeta$, which is approximately equal to the nondimensional height. 
}
\label{fig:orderparams}
\end{figure*}

In order to investigate the arrangements of neighboring particles, we calculated the height dependence of the directional order parameters. 

We regard two particles as a neighboring pair if the distance between their centers, $r^{(ij)}\equiv |\bm{r}^{(ij)}|$, is less than $35\;\mathrm{\mu m}$, where $\bm{r}^{(ij)}\equiv \bm{r}_j-\bm{r}_i$, is the vector pointing from the center of the $i$th particle to the $j$th particle.  
Assuming a height interval of thickness $60 l_p=104.4\;\mathrm{\mu m}$ on either side of $z(\zeta)$, we define the parameters 
 \begin{align}
 S_{kl}(\zeta)\equiv \langle n^{(ij)}_k n^{(ij)}_l-\frac{1}{3}\delta_{kl}\rangle_{\zeta}\quad  (k,l=x,y,z), \label{S}
\end{align}
where $\bm{n}^{(ij)}=(n^{(ij)}_x, n^{(ij)}_y, n^{(ij)}_z)\equiv \bm{r}^{(ij)}/r^{(ij)}$ and the angular brackets represent the average over all pairs of neighboring particles such that the $i$th particle is located within this interval. 
In Ref.~\cite{Kitsunezaki17} we investigated these order parameters for the entire region corresponding to the field of view and found that the anisotropy induced by shaking is reflected by the diagonal components. 

In this study, we investigate the quantities 
\begin{align}
S_1\equiv \frac{1}{\sqrt{2}}(S_{yy}-S_{xx}),\quad S_2\equiv \sqrt{\frac{3}{2}}(S_{yy}+S_{xx}), \nonumber\\
\mbox{and}\  S_\sigma^2\equiv \frac{2}{3}(S_{yz}^2+S_{zx}^2+S_{xy}^2). \label{S_1,S_2,S_sigma}
\end{align} 
We can regard $S_1$ and $S_2$ as anisotropy indices in the horizontal plane and vertical direction, respectively. 
The quantity $S_1$ increases as the number of neighboring particles increases in the direction perpendicular to that of the initial shaking in the shear plane, and $S_2$ decreases as the number increases in the vertical direction, as seen from the relation $S_{xx}+S_{yy}+S_{zz}=0$. 
The quantity $S_\sigma$ decreases as $S$ becomes closer to a diagonal matrix. 
As explained in the Appendix, the averages of $S_1$ and $S_2$ over samples would vanish and their standard deviations would be equal to $S_\sigma$ if every $\bm{n}^{(ij)}$ was chosen independently from an isotropic uniform distribution. 

Figure~\ref{fig:orderparams} plots the height dependences of $S_1$, $S_2$, and $S_\sigma$. 
Figure~\ref{fig:orderparams}(a) displays the results for two typical sets of conditions under which the memory effect of shaking appears, $(f,\phi(0))=(60\;\mathrm{rpm}, 18.4\%)$ and $(80\;\mathrm{rpm}, 18.4\%)$. 
Near the bottom of a layer, $S_1$ takes large positive values, while $S_2$ remains at zero, within the experimental uncertainty. 
This anisotropy is clearer for $f=80\;\mathrm{rpm}$ than for $f=60\;\mathrm{rpm}$. 
Although $\phi_g$ differed among the five samples prepared under each set of conditions in the range from approximately $\phi(0)$ to the value at which cracks appear, there was no significant difference in the degree of anisotropy among the samples.  
This indicates that the anisotropy is created initially and retained during desiccation. 

The observed height dependence is consistent with the fact that the memory effect of shaking is caused by shear stresses larger than the yield stress. Because the lower region of the paste experienced larger shear stresses during shaking, due to the weight of the upper region, the lower region was fluidized repeatedly, while the deformation of the upper region was mainly elastic and small. 
Although the nonlinear response to shaking is not simple, we infer that the memory of shaking is preserved mainly in the lower region, which experienced large deformation \cite{Goehring15}.

Figure~\ref{fig:orderparams}(b) displays the results for two sets of conditions under which the paste does not exhibit memory of shaking, $(f,\phi(0))=(0,18.4\%)$ and $(60\;\mathrm{rpm},10.4\%)$. 
Unlike the results displayed in Fig.~\ref{fig:orderparams}(a), these results do not exhibit anisotropy characterized by $S_1>S_2\simeq 0$, as expected. 
However, for both sets of conditions considered here, the arrangement of particles was not isotropic, with $S_2<0$. 
This anisotropy implies that the number of neighboring particles increases in the $z$ direction. 
In addition, there is a difference in $S_1$ between the two cases: $S_1$ is positive for the shaken samples with a small initial volume fraction of $\phi(0)=10.4\%$, while $S_1\simeq 0$ for the unshaken samples ($f=0$), although a large value of $S_\sigma$ indicates that the matrix $S$ is not fully diagonalized.
Such anisotropy is regarded as a kind of ``hidden memory'', which is not manifested as an anisotropic crack pattern. 

\section{Anisotropic structures of interstices\label{sec:interstices}}

\begin{figure*}[t]
\includegraphics[width=12.0cm]{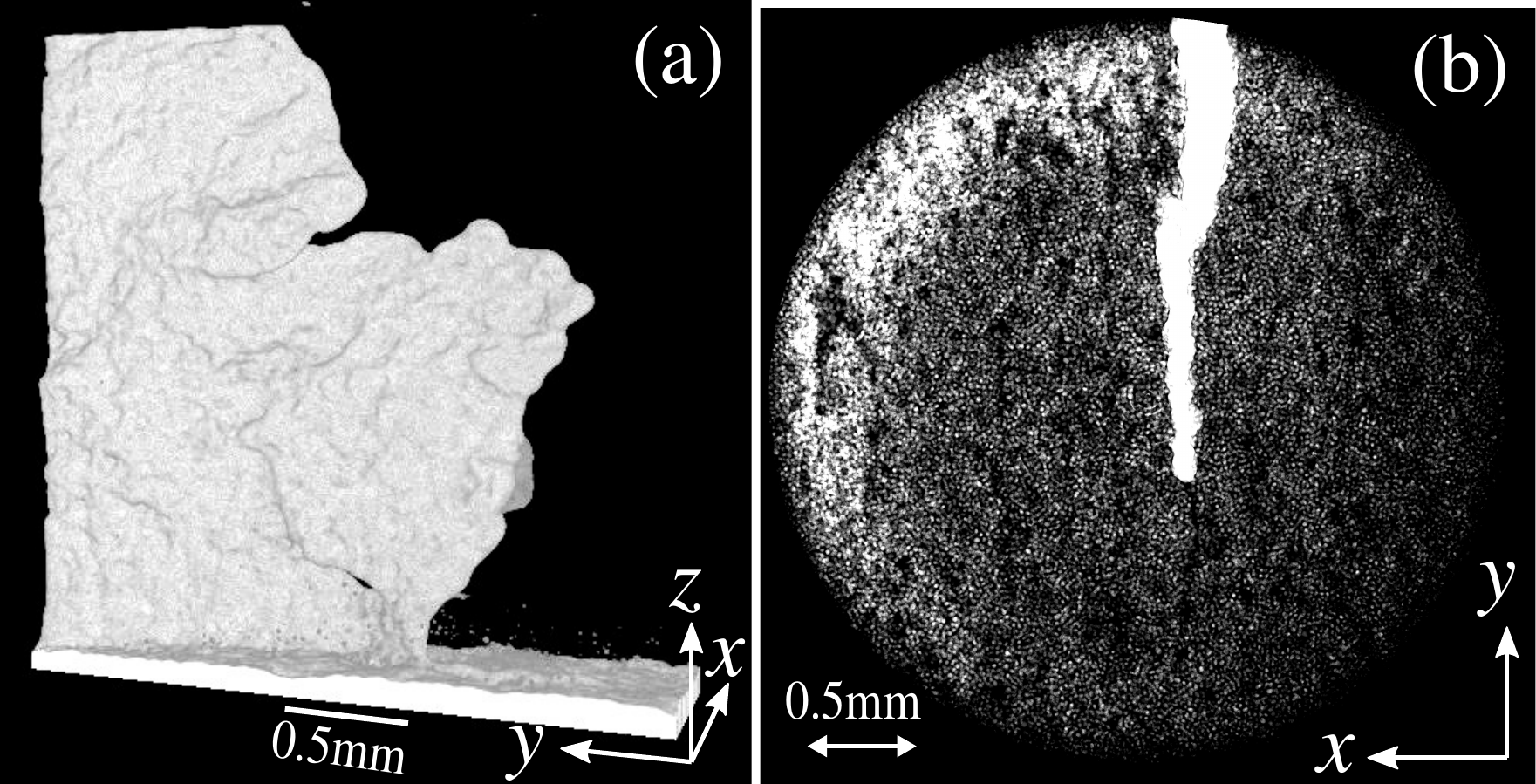} 
\caption{
$\mu$CT images of a piece of the layer. 
We cut the same sample as in Fig.~\ref{fig:cross_section} to include a crack tip. 
(a) Reconstructed 3D image of an air region penetrating into the layer of paste with crack growth. 
The air region appears white. 
As the sample was positioned upside down on the jig during the $\mu$CT scanning,  the layer of paste appears to be situated above the region of air. 
(b) Cross section at height $\zeta=0.5$. 
This image was obtained by averaging the brightnesses of 80 individual images, as in Fig.\ref{fig:cross_section}(c). 
}
\label{fig:cracktip}
\end{figure*}

We next investigate the interstices depicted in Figs.~\ref{fig:cross_section}(b) and \ref{fig:cross_section}(c). 
These interstices are retained during desiccation and significantly affect crack growth.

Because the sample considered in Fig.~\ref{fig:cross_section} was covered and solidified after several cracks had grown partway through the sample along the direction perpendicular to the initial shaking, these cracks were preserved. 
We cut out a piece of the layer including a crack tip and observed it with $\mu$CT. 
Figure~\ref{fig:cracktip}(a) is a 3D image of the air regions constructed from the $\mu$CT data \cite{3DDistanceMap}. 
In this image, the growing crack appears as the penetration of air into the paste. 
The crack tip has a complex shape, which suggests unstable growth. 
It is also seen that the region of intermediate height leads the growth of the crack, with the lower and upper regions trailing behind.  
These characteristics of the crack growth are consistent with what we have inferred from plumose patterns left on crack surfaces in desiccation cracks, although we have not observed a sharp plumose structure in Lycopodium paste \cite{Weinberger99, Weinberger01, Kitsunezaki09}. 
In this work we studied four crack tips cut from three samples. 
Two of them had similar properties, while for the other two, the tip shapes deviated to the top or bottom. 

Figure~\ref{fig:cracktip}(b) displays a horizontal cross section at the center of the layer considered in Fig.~\ref{fig:cracktip}(a). 
Here a running average in the depth direction was carried out in order to make the interstices clearly visible, as in Fig.~\ref{fig:cross_section}(c). 
Note that there is an interstice ahead of the crack tip, and the crack runs along this interstice in the direction perpendicular to that of the initial shaking, although the width of the interstice is extended inside the crack. 
Similar relations between the interstices and crack growth were observed in all four crack tips. 
We thus infer that the structures of the interstices determine the direction of crack growth, and for this reason, they are directly responsible for the memory effect of shaking exhibited by Lycopodium paste. 

\begin{figure*}[t]
\includegraphics[width=16.0cm]{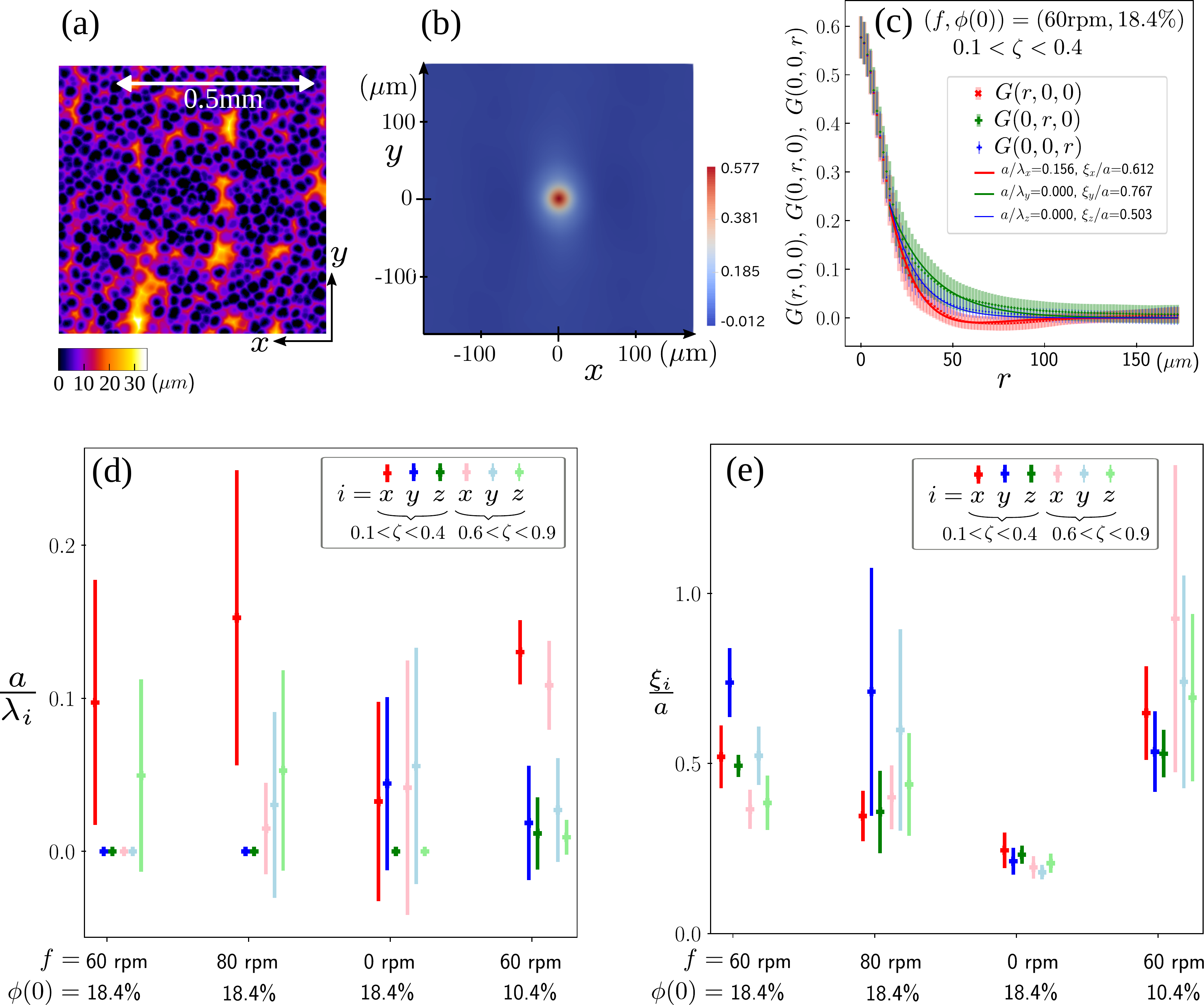} 
\caption{
Correlation functions $G(\bm{r})$ of distance maps: 
(a) typical horizontal cross section of a distance map, (b) image of $G(x,y,0)$, and (c) plots of $G(\bm{r})$ in the $x,y$, and $z$ directions, all for the lower part of the sample considered in Fig.~\ref{fig:cross_section}. 
The image in (a) depicts the central region at the same height as in Fig.~\ref{fig:cross_section}(c),  
(b) is the average of $G(x,y,0)$ over $72$ cubic regions with side lengths $200l_p$, and  
(c) plots the averages of $G(r,0,0)$, $G(0,r,0)$, and $G(0,0,r)$ with fitting functions proportional to $\cos(2\pi \frac{r}{\lambda_i})e^{-r/\xi_i}$ ($i=x,y,z$). 
Also shown are plots of the averages of (d) $a/\lambda_i$ and (e) $\xi_i/a$ over five samples prepared under each set of conditions indicated on the horizontal axis. 
Here $a=30\;\mathrm{\mu m}$ is a typical particle diameter. 
The data are arranged in the order $x,y,z$ for the lower part, followed by $x,y,z$ for the upper part, and the error bars indicate the standard deviations.
}
\label{fig:distacemap}
\end{figure*}

In order to ascertain the properties of interstices from a 3D binary image of particles, we constructed a 3D distance map $D(\bm{r})$, which represents the distance from each position $\bm{r}$ to the nearest particle [see Fig.~\ref{fig:distacemap}(a)], and calculated the correlation functions using the equation  
\begin{align}
G(\bm{r})\equiv \frac{\frac{1}{V}\int d^3\bm{r}'D(\bm{r}')D(\bm{r}'+\bm{r})}{\left(\frac{1}{V}\int d^3\bm{r}'D(\bm{r}')\right)^2}-1.
\end{align}
We divided both the lower part ($0.1<\zeta<0.4$) and the upper part ($0.6<\zeta<0.9$) of a sample into cubic regions with side lengths $200 l_p=348\;\mathrm{\mu m}$ to calculate $G(\bm{r})$ in every region, except several regions that contained large impurities, such as bubbles. 

The function $G(\bm{r})$ reflects the statistical properties of the shape of an interstice.  
For the lower part of the sample considered in Fig.~\ref{fig:cross_section}, the average of $G(x,y,0)$ over all regions is not isotropic, as seen in Fig.~\ref{fig:distacemap}(b). 
Figure~\ref{fig:distacemap}(c) plots $G(r,0,0)$, $G(0,r,0)$, and $G(0,0,r)$ with their fitting functions. 
It is seen that $G(\bm{r})$ decays monotonically in the $y$ and $z$ directions, while it decays faster and has a minimum in the $x$ direction. 
This difference implies that interstices have anisotropic shapes that are shorter in the $x$ direction. 
Defining $a\equiv 30\;\mathrm{\mu m}$ as a typical diameter of a Lycopodium particle, we fitted $G(r,0,0)$, $G(0,r,0)$, and $G(0,0,r)$ for $r>\frac{a}{2}$ with functions proportional to $e^{-r/\xi_i}\cos{\frac{2\pi r}{\lambda_i}}$ ($i=x,y,z$), respectively, where the fitting parameter $1/\lambda_i$ vanishes for the monotonically decreasing functions. 
The fitting parameter $\xi_i$ represents the correlation length in the $i$th direction and $\lambda_i/2$ corresponds to a typical distance from the inside of the interstices to the high-density regions in the case that $G$ has a minimum. 
We note that these fitting functions are adopted as the first approximation, as $G(\bm{r})$ does not have such a simple form; specifically, it tends to decay somewhat more slowly than an exponential function in the $y$ direction. 

We calculated $G(\bm{r})$ for the lower and upper parts of the sets of samples considered in Fig.~\ref{fig:orderparams}.  
The averages of $a/\lambda_i$ and $\xi_i/a$ over five samples for each set of conditions are plotted in Figs.~\ref{fig:distacemap}(d) and \ref{fig:distacemap}(e), respectively. 
For the two sets of conditions under which the memory effect of shaking appears, we find distinct anisotropy in the lower parts of the samples: $a/\lambda_y$ and $a/\lambda_z$ approximately vanish and $a/\lambda_x$ is approximately $0.1$ for $(f,\phi(0))=(60\;\mathrm{rpm}, 18.4\%)$ and $0.15$ for $(80\;\mathrm{rpm}, 18.4\%)$, while $\xi_y$ tends to be larger than $\xi_x$ and $\xi_z$.  
We thus conclude that anisotropic interstices develop perpendicularly to the $x$ direction in the lower part of a sample, with typical widths in the range $\lambda_x/2=3a$--$5a$ in the $x$ direction, while they extend mainly in the $y$ direction. 
Contrastingly, such anisotropy is not found in the upper part under the same conditions, although $a/\lambda_z$ tends to be large. 

For the two sets of conditions under which the memory effect of shaking does not appear, the forms of $G(\bm{r})$ differ significantly. 
In the unshaken case $(0\;\mathrm{rpm}, 18.4\%)$, we found no anisotropy in the horizontal plane, as expected, and $\xi_x$, $\xi_y$, and $\xi_z$ have similar values that are smaller than those in the other cases. 
This result suggests that interstices develop through repeated shear deformation. 
For the shaken samples with small initial solid volume fractions $(60\;\mathrm{rpm}, 10.4\%)$, we found anisotropy with $a/\lambda_x$ taking large values in both the lower and upper parts. 
Also in this case, large density fluctuations appeared in the form of voids rather than anisotropic interstices in the 3D images. 
Because paste is less viscous for small solid volume fractions, we infer that shaking causes this anisotropy with large shear flows, but such anisotropy is not reflected in crack patterns. 

These results indicate that anisotropic structures of interstices are created by shaking in paste with a large solid volume fraction, just as for the anisotropic arrangements of neighboring particles reported in Sec.~\ref{sec:orderparams}. 

\section{Numerical Simulations\label{sec:discussion}}

With the experimental conditions used in this work, the Lycopodium particles in paste can be regarded as non-Brownian particles in high-viscosity shear flow. 
The particle Reynolds number and the P\'eclet number are estimated as $\mathrm{Re}_p\equiv \dot{\gamma}a^2 \rho_w/\eta_w \simeq 10^{-3}\ll 1$ and $\mathrm{Pe}\equiv 6\pi \eta_w \dot{\gamma}a^3/(k_BT)\simeq 10^5\gg 1$, respectively, for Lycopodium particles of diameter $a=3\times 10^{-5}\;\mathrm{m}$ \cite{Denn18}. 
Here, $k_B$ is the Boltzmann constant, and $\eta_w$ and $\rho_w$ are the viscosity and density of water, for which we used the values $\eta_w\simeq 7\times 10^{-4}\;\mathrm{Pa\;s}$ and $\rho_w\simeq 10^{3}\;\mathrm{kg/m^3}$. 
We used a shear rate of $\dot{\gamma}\simeq 1\;\mathrm{s^{-1}}$ and an absolute temperature of $T\simeq 3\times 10^2\;\mathrm{K}$ as typical experimental conditions. 
Some previous works investigated rearrangements of non-Brownian particles in Stokes flows \cite{Roche11,Denn18,Guazzelli18,Varga19}. 
It is known that the motion of such particles is not reversible under oscillating shear flow.   
In particular, it was reported that colloidal particles confined between two parallel plates become arranged in the direction perpendicular to the flow direction under an oscillating shear flow \cite{Cheng12}. 

As a first step in investigating how shaking creates anisotropic arrangements, we performed numerical simulations of spherical particles under a given shear flow \cite{Olsson07}. 
We assumed that a uniform shear flow of given shear rate $\dot{\gamma}(t)$ exerts viscous drag on every particle and that this viscous drag always balances with the contact forces exerted by other particles, due to the large viscosity. 
In these simulations, we found that there appear anisotropic arrangements of neighboring particles similar to those reported in Sec.~\ref{sec:orderparams}, but large density fluctuations, such as those forming interstices, do not appear.    

We numerically integrated the equations 
\begin{align}
\dot{\bm{r}}_i=(\dot{\gamma}(t)z_i,0,0)+\frac{1}{2R_i}\sum_{j(\neq i)}f(r^{(ij)})\bm{n}^{(ij)} \label{dot_r_i}
\end{align}
to generate the time evolution of the positions of the $N$ particles, $\bm{r}_i=(x_i(t),y_i(t),z_i(t))$ \ $(i=1,2,...,N)$, where the time $t$ is merely a parameter used to define the shear deformation $\gamma(t)$. 
The first term in Eq.~(\ref{dot_r_i}) represents the velocity produced by a simple shear flow. 
We investigated two cases, that of oscillating shear flow, with $\gamma(t)=\gamma_m\sin{(2\pi t)}$, and that of constant shear flow, with $\gamma(t)=4\gamma_m t$, where the magnitudes of the shear strains were set to change by $\gamma_m$ per quarter of time in both cases. 
The second term in Eq.~(\ref{dot_r_i}) represents short-range elastic interactions between spherical particles in contact. 
The factor $1/2R_i$ comes from the size dependence of the mobility of a particle in Stokes flows. 
We assumed normal forces of Hertzian contacts in the form 
\begin{align}
 f(r^{(ij)})\equiv 
-\alpha\sqrt{\frac{R_iR_j}{R_i+R_j}}(R_i+R_j-r^{(ij)})^{3/2} 
\end{align}
for $R_i+R_j>r^{(ij)}$ and $f(r^{(ij)})=0$ otherwise, where $R_i$ and $R_j$ are the radii of the $i$th and $j$th particles. 
In other words, particles interact repulsively when they overlap. 
We determined the value of $\alpha$ to be $100$ from numerical simulations in order to maintain an average overlap of less than $5\%$. 

Taking the average diameter of a particle as the unit of length, the particle sizes were uniformly distributed over the interval $[0.8,1.2]$ and the system size was $20$. 
The system was a cubic region with periodic boundary conditions in the $x$ and $y$ directions and Lees-Edwards boundary conditions in the $z$ direction. 
We distributed the $N$ particles randomly at the initial time $t=0$ using the method described in Ref.~\cite{Li08} and used the midpoint method with a time step $\mathit{\Delta}t=0.001$. 

\begin{figure*}[t]
\includegraphics[width=17.5cm]{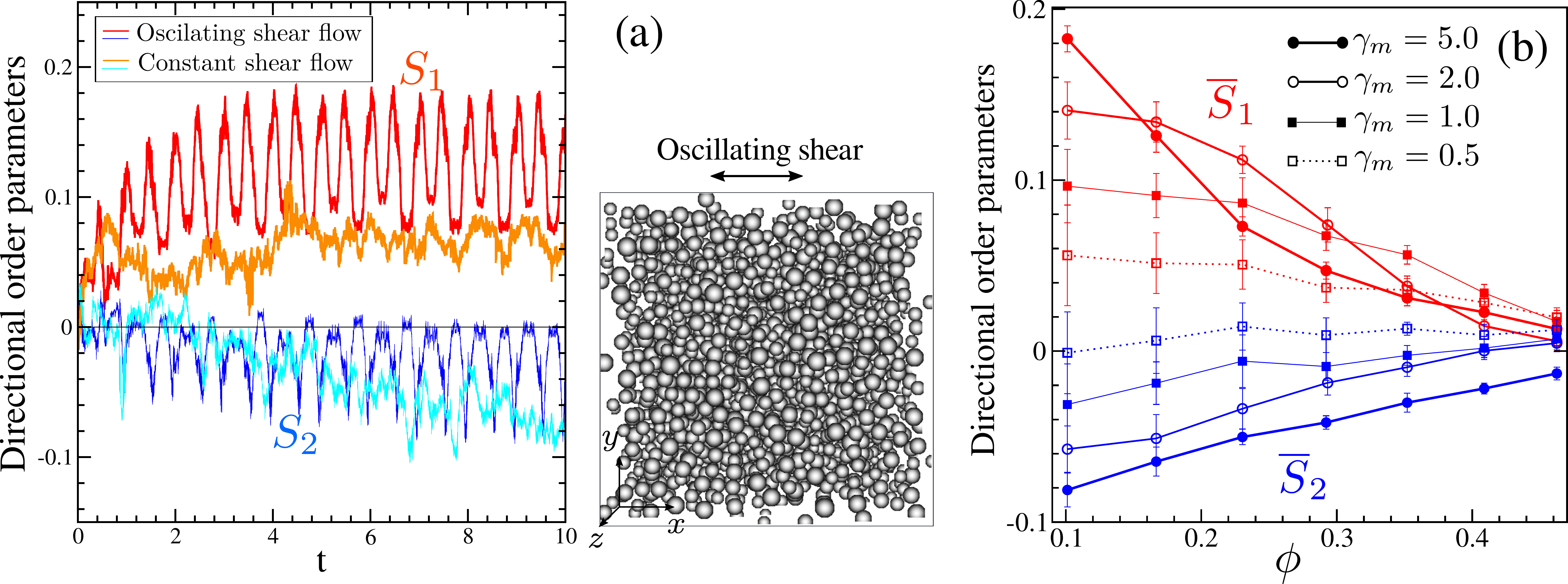} 
\caption{
(a) Directional order parameters obtained from numerical simulations in which $N=3500$ spherical particles  were subject to shear deformation beginning with a random initial arrangement in a cubic system and a typical snapshot of particles in the system after a sufficiently long period of oscillating shear flow ($t=20$). 
The time evolutions of $S_1$ and $S_2$ are plotted in the cases of oscillating shear flow and constant shear flow, respectively.   
(b) For the cases of oscillating shear flow, we investigated the time averages of the order parameters in the period $19.0\leq t<20.0$, $\overline{S}_1$ and $\overline{S}_2$,  using various values of $N$ and the amplitude $\gamma_m$.  The averages of $\overline{S}_1$ and $\overline{S}_2$ over $16$ initial conditions are plotted with respect to the volume fraction $\phi$. The error bars represent the standard deviations. 
\label{fig:motion_under_shear}}
\end{figure*}

Figure~\ref{fig:motion_under_shear}(a) plots the time evolutions of the order parameters obtained from the numerical simulations for a solid volume fraction of $\phi=22.8\%$, which is similar to the value of $\phi(0)$ used in our experiments. 
Here we chose $\gamma_m=2$ as the lower region of a sample is inferred to experience shear deformation of the order of 1 in experiments \cite{Goehring15}. 
The definitions of the order parameters are the same as in Sec.~\ref{sec:orderparams}, where two particles are regarded as a neighboring pair if the center-to-center distance is less than $1.16 \simeq 35\;\mathrm{\mu m}/a$. 
As seen in Fig.~\ref{fig:motion_under_shear}(a), in the case of oscillating shear flow, we find anisotropy, with $S_1>0$ and $S_2\simeq 0$. 
This is similar to the anisotropy depicted in Fig.~\ref{fig:orderparams}(a). 
The anisotropy emerges quickly within a few cycles of shaking and it is consistent with the recent experimental finding that the memory of shaking can be rewritten through the influence of one or two oscillations in a different direction \cite{Nakahra19,Morita21}. 
Figure~\ref{fig:motion_under_shear}(b) plots the dependence of this anisotropy on the volume fraction $\phi$ and the amplitude of shear deformation $\gamma_m$.  It is seen that the anisotropy becomes weak as $\gamma_m$ decreases or $\phi$ increases. We conclude that collisions among loosely packed particles under large oscillating deformation can yield short-range anisotropy. 

Figure~\ref{fig:motion_under_shear}(a) also indicates that, in the case of constant shear flow, another type of anisotropy, with $S_2<0<S_1$, appears as the shear deformation increases.   
This anisotropy is similar to that found for the shaken samples with small $\phi(0)$ considered in Fig.~\ref{fig:orderparams}(b). 
This is reasonable, because paste is fluidized entirely during shaking, due to the very small yield stress realized under such conditions. 
This should be compared to the case of the unshaken samples considered in Fig.~\ref{fig:orderparams}(b), for which $S_2<0$, $S_1\simeq 0$, and the matrix $S$ is not fully diagonalized.     
We infer that this latter type of anisotropy was caused by uncontrolled shear flows created when the paste was poured into the container initially. 
In this case, $S_1$ vanishes as a result of the average over the five samples, because the samples experienced flows in various directions. 

In contrast to the anisotropy in the arrangement of neighboring particles,
we were not able to generate structures of interstices similar to those seen experimentally in our numerical simulations. 
This can be attributed to the fact that the model used in this work is too simple to describe the rheological properties of Lycopodium paste, in particular, the yield stresses created with small solid volume fractions. 
Lycopodium particles exhibit interactions that are much more complicated than the simple repulsive interactions assumed in the simulations. 
As Lycopodium particles consist mainly of fatty oil and have porous surfaces, the surfaces become hydrophobic when the fine asperities trap air. 
However, recent experiments investigating water droplets on a surface composed of Lycopodium particles found that vertical vibration induces a wetting transition through which the surface becomes hydrophilic \cite{Bormashenko09,Bormashenko09b}. 
It is likely that there is a similar transition that removes air from the porous surfaces when a mixture is stirred in the preparation of a paste. 
This can be understood from the fact that surfaces experience large stresses during stirring, just as in the case of vibration. 
After such processes, it is likely that adhesive and frictional forces are exerted between the porous surfaces of Lycopodium particles in contact. 
Studies of jamming transitions have found that adding attractive forces or frictional forces to rigid particles reduces the jamming point so that yield stresses emerge at smaller volume fractions \cite{Seto13,Vinutha16,Koeze20}. 
From these considerations, we conjecture that there is a sparse network of particles that supports yield stresses, and shaking could make such a network structure anisotropic. 
Our experimental results suggest that such a structure develops irreversibly under oscillating shear deformation if the suspension of non-Brownian particles behaves as a plastic fluid.

\section{Conclusion\label{sec:conclusion}}

We carried out $\mu$CT observations of the 3D arrangements of particles in Lycopodium paste to elucidate the structures responsible for the memory effect of shaking. 
We found that applying horizontal shaking in one direction induces anisotropic structures mainly in the lower part of a layer of paste; 
the number of neighboring particles increases in the direction perpendicular to the shaking in the horizontal plane, and density fluctuations also emerge as anisotropic interstices extending in the perpendicular direction. 
Numerical simulations of non-Brownian particles under a given shear flow indicate that collisions of particles can account for the anisotropic arrangements of neighboring particles. 
We conclude that the formation of anisotropic interstices is directly responsible for the memory effect of shaking exhibited by Lycopodium paste.
Interstices are robust during desiccation and they play a role as a path of air penetration causing anisotropic crack growth in the direction parallel to the initial shaking. 
We do not yet understand the process through which anisotropic interstices are created nor yield stresses in the case of small solid volume fractions.  
We leave for future work the investigate of how yield stresses emerge in systems with low particle densities and how shaking creates anisotropic interstices in such systems.

\appendix* 

\section{Order parameters for an isotropic distribution
\label{sec: Orderparameters for  the isotropic distributuon} }

\begin{figure}[t]
\includegraphics[width=8cm]{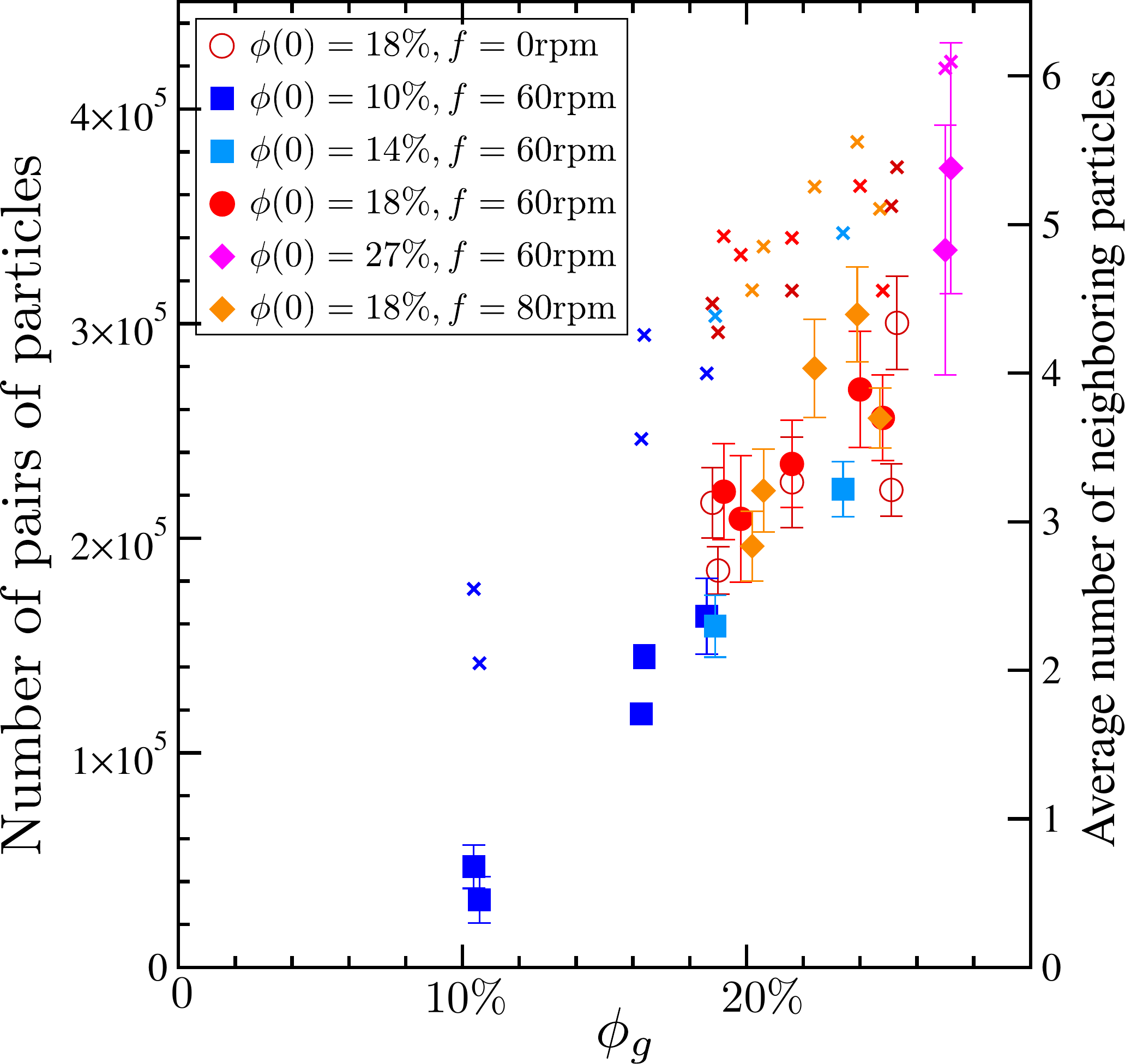} 
\caption{
Numbers of pairs of neighboring particles existing in a height interval of thickness $2\times 60l_p$. 
These values were used to calculate the data points in Fig.~\ref{fig:orderparams}. 
The average over all such intervals in each sample is plotted along with the standard deviation. 
The crosses indicate the average numbers of neighboring particles per particle with respect to the axis on the right-hand side.}
\label{fig:M}
\end{figure}

The values of $S_i$ in Fig.~\ref{fig:orderparams} have fluctuations resulting from the finiteness of the number of unit vectors used in the calculations.   
When $S_{kl}\equiv \frac{1}{M}\sum_{m=1}^Mn^{(m)}_kn^{(m)}_l-\frac{1}{3}\delta_{kl}$ is calculated using $M$ unit vectors, where $\bm{n}^{(m)}$ is the unit vector pointing from the center of one particle to the center of the other in the $m$th pair of particles,  
if $\{\bm{n}^{(m)}\}$ were generated from an isotropic distribution independently, the average and variation of $S_{kl}$ with respect to the probability distribution would be 
\begin{align}
\overline{S_{kl}}=0,
\quad 
\overline{S_{ij}S_{kl}}= \frac{1}{15M}\left(-\frac{2}{3}\delta_{ij}\delta_{kl}+\delta_{ik}\delta_{jl}
+\delta_{il}\delta_{jk}\right), 
\end{align}
respectively, since $\overline{n_i n_j}=\frac{1}{3}\delta_{ij}$ and $\overline{n_i n_j n_k n_l}=\frac{1}{15}(\delta_{ij}\delta_{kl}+\delta_{ik}\delta_{jl}+\delta_{il}\delta_{jk})$. 

Because Eq.~(\ref{S}) defines a symmetric matrix $S$ with the constraint $\mathrm{Tr}S\equiv S_{xx}+S_{yy}+S_{zz}=0$, the number of independent components is five and there is correlation among the diagonal components. 
Defining $S_3\equiv \sqrt{2}S_{yz}$, $S_4\equiv \sqrt{2}S_{zx}$ and $S_5\equiv \sqrt{2}S_{xy}$, in addition to $S_1$ and $S_2$ defined in Eq.(\ref{S_1,S_2,S_sigma}), we have 
\begin{align}
\overline{S}_i=0,\quad \overline{S_iS_j}=\frac{2}{15M}\delta_{ij}\quad
(i,j=1,2,...,5), 
\end{align}
with $\overline{\mathrm{Tr}S^2}=\sum_{i=1}^5 \overline{S_i^2}$ and $S_\sigma^2=\frac{1}{3}(S_3^2+S_4^2+S_5^2)$.

Figure~\ref{fig:M} plots the number of pairs of neighboring particles used to calculate each data point in Fig.~\ref{fig:orderparams}.   
We find that this number is determined approximately by the solid volume fraction at the gelation time, $\phi_g$.
Here we counted the pairs $(i,j)$ and $(j,i)$ separately, and thus $M$ is approximately half of the indicated number of pairs.  
Because $2M=(2\times 10^4)-(4\times 10^5)$, the standard deviation of $S_i$ is expected to be $\sqrt{\frac{2}{15M}}=(1-4)\times 10^{-3}$ for an isotropic distribution.

\begin{acknowledgments}
We thank K.~Uesugi and M.~Hoshino, beam line scientists at SPring-8. 
We also thank M.~Otsuki and Ooshida T. for discussions of theoretical topics and G. C. Paquette for valuable comments.  
Our $\mu$CT observations were performed at the BL20B2 of SPring-8 with the approval of the Japan Synchrotron Radiation Research Institute (Proposals No. 2018B1477 and No. 2019B1150). 
This research was supported by two Grants-in-Aid for Scientific Research (Grants No. KAKENHI C 18K03560
 and No. KAKENHI C 20K03886) from JSPS, Japan. 
\end{acknowledgments}



\begin{thebibliography}{33}%
\makeatletter
\providecommand \@ifxundefined [1]{%
 \@ifx{#1\undefined}
}%
\providecommand \@ifnum [1]{%
 \ifnum #1\expandafter \@firstoftwo
 \else \expandafter \@secondoftwo
 \fi
}%
\providecommand \@ifx [1]{%
 \ifx #1\expandafter \@firstoftwo
 \else \expandafter \@secondoftwo
 \fi
}%
\providecommand \natexlab [1]{#1}%
\providecommand \enquote  [1]{``#1''}%
\providecommand \bibnamefont  [1]{#1}%
\providecommand \bibfnamefont [1]{#1}%
\providecommand \citenamefont [1]{#1}%
\providecommand \href@noop [0]{\@secondoftwo}%
\providecommand \href [0]{\begingroup \@sanitize@url \@href}%
\providecommand \@href[1]{\@@startlink{#1}\@@href}%
\providecommand \@@href[1]{\endgroup#1\@@endlink}%
\providecommand \@sanitize@url [0]{\catcode `\\12\catcode `\$12\catcode
  `\&12\catcode `\#12\catcode `\^12\catcode `\_12\catcode `\%12\relax}%
\providecommand \@@startlink[1]{}%
\providecommand \@@endlink[0]{}%
\providecommand \url  [0]{\begingroup\@sanitize@url \@url }%
\providecommand \@url [1]{\endgroup\@href {#1}{\urlprefix }}%
\providecommand \urlprefix  [0]{URL }%
\providecommand \Eprint [0]{\href }%
\providecommand \doibase [0]{https://doi.org/}%
\providecommand \selectlanguage [0]{\@gobble}%
\providecommand \bibinfo  [0]{\@secondoftwo}%
\providecommand \bibfield  [0]{\@secondoftwo}%
\providecommand \translation [1]{[#1]}%
\providecommand \BibitemOpen [0]{}%
\providecommand \bibitemStop [0]{}%
\providecommand \bibitemNoStop [0]{.\EOS\space}%
\providecommand \EOS [0]{\spacefactor3000\relax}%
\providecommand \BibitemShut  [1]{\csname bibitem#1\endcsname}%
\let\auto@bib@innerbib\@empty
\bibitem [{\citenamefont {Coussot}(2005)}]{Coussot05}%
  \BibitemOpen
  \bibfield  {author} {\bibinfo {author} {\bibfnamefont {P.}~\bibnamefont
  {Coussot}},\ }\href@noop {} {\emph {\bibinfo {title} {Rheometry of Pastes,
  Suspensions, and Granular Materials}}}\ (\bibinfo  {publisher} {Wiley},\ \bibinfo {address} {Hoboken},\ \bibinfo {year}
  {2005})\BibitemShut {NoStop}%
\bibitem [{\citenamefont {Goehring}\ \emph {et~al.}(2015)\citenamefont
  {Goehring}, \citenamefont {Nakahara}, \citenamefont {Dutta}, \citenamefont
  {Kitsunezaki},\ and\ \citenamefont {Tarafdar}}]{Goehring15}%
  \BibitemOpen
  \bibfield  {author} {\bibinfo {author} {\bibfnamefont {L.}~\bibnamefont
  {Goehring}}, \bibinfo {author} {\bibfnamefont {A.}~\bibnamefont {Nakahara}},
  \bibinfo {author} {\bibfnamefont {T.}~\bibnamefont {Dutta}}, \bibinfo
  {author} {\bibfnamefont {S.}~\bibnamefont {Kitsunezaki}},\ and\ \bibinfo
  {author} {\bibfnamefont {S.}~\bibnamefont {Tarafdar}},\ }\href@noop {} {\emph
  {\bibinfo {title} {Desiccation Cracks and their Patterns: Formation and
  Modelling in Science and Nature}}} (\bibinfo  {publisher} {Wiley},\ New York, \bibinfo {year}
  {2015})\BibitemShut {NoStop}%
\bibitem [{\citenamefont {Nakahara}\ and\ \citenamefont
  {Matsuo}(2005)}]{Nakahara05}%
  \BibitemOpen
  \bibfield  {author} {\bibinfo {author} {\bibfnamefont {A.}~\bibnamefont
  {Nakahara}}\ and\ \bibinfo {author} {\bibfnamefont {Y.}~\bibnamefont
  {Matsuo}},\ }\href@noop {} {\bibfield  {journal} {\bibinfo  {journal} {J.
  Phys. Soc. Jpn.}\ }\textbf {\bibinfo {volume} {74}},\ \bibinfo {pages} {1362}
  (\bibinfo {year} {2005})}\BibitemShut {NoStop}%
\bibitem [{\citenamefont {Nakahara}\ and\ \citenamefont
  {Matsuo}(2006{\natexlab{a}})}]{Nakahara06}%
  \BibitemOpen
  \bibfield  {author} {\bibinfo {author} {\bibfnamefont {A.}~\bibnamefont
  {Nakahara}}\ and\ \bibinfo {author} {\bibfnamefont {Y.}~\bibnamefont
  {Matsuo}},\ }\href@noop {} {\bibfield  {journal} {\bibinfo  {journal} {Phys.
  Rev. E}\ }\textbf {\bibinfo {volume} {74}},\ \bibinfo {pages} {045102(R)}
  (\bibinfo {year} {2006}{\natexlab{a}})}\BibitemShut {NoStop}%
\bibitem [{\citenamefont {Nakahara}\ and\ \citenamefont
  {Matsuo}(2006{\natexlab{b}})}]{Nakahara06b}%
  \BibitemOpen
  \bibfield  {author} {\bibinfo {author} {\bibfnamefont {A.}~\bibnamefont
  {Nakahara}}\ and\ \bibinfo {author} {\bibfnamefont {Y.}~\bibnamefont
  {Matsuo}},\ }\href@noop {} {\bibinfo  {journal} {J. Stat. Mech.: Theory
  Exp.}\ (\bibinfo {year} {2006})} \bibinfo {pages} {P07016}\BibitemShut {NoStop}%
\bibitem [{\citenamefont {Nakahara}\ \emph {et~al.}(2011)\citenamefont
  {Nakahara}, \citenamefont {Shinohara},\ and\ \citenamefont
  {Matsuo}}]{Nakahara11}%
  \BibitemOpen
\bibfield  {journal} {  }\bibfield  {author} {\bibinfo {author} {\bibfnamefont
  {A.}~\bibnamefont {Nakahara}}, \bibinfo {author} {\bibfnamefont
  {Y.}~\bibnamefont {Shinohara}},\ and\ \bibinfo {author} {\bibfnamefont
  {Y.}~\bibnamefont {Matsuo}},\ }\href@noop {} {\bibfield  {journal} {\bibinfo
  {journal} {J. Phys. : Conf. Ser.}\ }\textbf {\bibinfo {volume} {319}},\
  \bibinfo {pages} {012014} (\bibinfo {year} {2011})}\BibitemShut {NoStop}%
\bibitem [{\citenamefont {Matsuo}\ and\ \citenamefont
  {Nakahara}(2012)}]{Matsuo12}%
  \BibitemOpen
  \bibfield  {author} {\bibinfo {author} {\bibfnamefont {Y.}~\bibnamefont
  {Matsuo}}\ and\ \bibinfo {author} {\bibfnamefont {A.}~\bibnamefont
  {Nakahara}},\ }\href@noop {} {\bibfield  {journal} {\bibinfo  {journal} {J.
  Phys. Soc. Jpn.}\ }\textbf {\bibinfo {volume} {81}},\ \bibinfo {pages}
  {024801} (\bibinfo {year} {2012})}\BibitemShut {NoStop}%
\bibitem [{\citenamefont {Nakayama}\ \emph {et~al.}(2013)\citenamefont
  {Nakayama}, \citenamefont {Matsuo}, \citenamefont {Takeshi},\ and\
  \citenamefont {Nakahara}}]{Nakayama13}%
  \BibitemOpen
  \bibfield  {author} {\bibinfo {author} {\bibfnamefont {H.}~\bibnamefont
  {Nakayama}}, \bibinfo {author} {\bibfnamefont {Y.}~\bibnamefont {Matsuo}},
  \bibinfo {author} {\bibfnamefont {Ooshida}~\bibnamefont {T.}},\ and\ \bibinfo
  {author} {\bibfnamefont {A.}~\bibnamefont {Nakahara}},\ }\href@noop {}
  {\bibfield  {journal} {\bibinfo  {journal} {Eur. Phys. J. E}\ }\textbf
  {\bibinfo {volume} {36}},\ \bibinfo {pages} {1} (\bibinfo {year}
  {2013})}\BibitemShut {NoStop}%
\bibitem [{\citenamefont {Kitsunezaki}\ \emph {et~al.}(2017)\citenamefont
  {Kitsunezaki}, \citenamefont {Sasaki}, \citenamefont {Nishimoto},
  \citenamefont {Mizuguchi}, \citenamefont {Matsuo},\ and\ \citenamefont
  {Nakahara}}]{Kitsunezaki17}%
  \BibitemOpen
  \bibfield  {author} {\bibinfo {author} {\bibfnamefont {S.}~\bibnamefont
  {Kitsunezaki}}, \bibinfo {author} {\bibfnamefont {A.}~\bibnamefont {Sasaki}},
  \bibinfo {author} {\bibfnamefont {A.}~\bibnamefont {Nishimoto}}, \bibinfo
  {author} {\bibfnamefont {T.}~\bibnamefont {Mizuguchi}}, \bibinfo {author}
  {\bibfnamefont {Y.}~\bibnamefont {Matsuo}},\ and\ \bibinfo {author}
  {\bibfnamefont {A.}~\bibnamefont {Nakahara}},\ }\href@noop {} {\bibfield
  {journal} {\bibinfo  {journal} {Eur. Phys. J. E}\ }\textbf
  {\bibinfo {volume} {40}},\ \bibinfo {pages} {88} (\bibinfo {year}
  {2017})}\BibitemShut {NoStop}%
\bibitem [{\citenamefont {Nakahara}\ \emph {et~al.}(2019)\citenamefont
  {Nakahara}, \citenamefont {Hiraoka}, \citenamefont {Hayashi}, \citenamefont
  {Matsuo},\ and\ \citenamefont {Kitsunezaki}}]{Nakahra19}%
  \BibitemOpen
  \bibfield  {author} {\bibinfo {author} {\bibfnamefont {A.}~\bibnamefont
  {Nakahara}}, \bibinfo {author} {\bibfnamefont {T.}~\bibnamefont {Hiraoka}},
  \bibinfo {author} {\bibfnamefont {R.}~\bibnamefont {Hayashi}}, \bibinfo
  {author} {\bibfnamefont {Y.}~\bibnamefont {Matsuo}},\ and\ \bibinfo {author}
  {\bibfnamefont {S.}~\bibnamefont {Kitsunezaki}},\ }\href@noop {} {\bibfield
  {journal} {\bibinfo  {journal} {Philos. Trans. R. Soc. A}\ }\textbf {\bibinfo {volume} {377}},\ \bibinfo {pages} {20170395}
  (\bibinfo {year} {2019})}\BibitemShut {NoStop}%
\bibitem [{\citenamefont {Otsuki}(2005)}]{Otsuki05}%
  \BibitemOpen
  \bibfield  {author} {\bibinfo {author} {\bibfnamefont {M.}~\bibnamefont
  {Otsuki}},\ }\href@noop {} {\bibfield  {journal} {\bibinfo  {journal} {Phys.
  Rev. E}\ }\textbf {\bibinfo {volume} {72}},\ \bibinfo {pages} {046115}
  (\bibinfo {year} {2005})}\BibitemShut {NoStop}%
\bibitem [{\citenamefont {Takeshi}(2008)}]{Ooshida08}%
  \BibitemOpen
  \bibfield  {author} {\bibinfo {author} {\bibfnamefont {Ooshida}~\bibnamefont
  {T.}},\ }\href@noop {} {\bibfield  {journal} {\bibinfo  {journal} {Phys.
  Rev. E}\ }\textbf {\bibinfo {volume} {77}},\ \bibinfo {pages} {061501}
  (\bibinfo {year} {2008})}\BibitemShut {NoStop}%
\bibitem [{\citenamefont {Takeshi}(2009)}]{Ooshida09}%
  \BibitemOpen
  \bibfield  {author} {\bibinfo {author} {\bibfnamefont {Ooshida}~\bibnamefont
  {T.}},\ }\href@noop {} {\bibfield  {journal} {\bibinfo  {journal} {J.
  Phys. Soc. Jpn.}\ }\textbf {\bibinfo {volume} {78}},\ \bibinfo {pages}
  {104801} (\bibinfo {year} {2009})}\BibitemShut {NoStop}%
\bibitem [{\citenamefont {Kitsunezaki}\ \emph {et~al.}(2016)\citenamefont
  {Kitsunezaki}, \citenamefont {Nakahara},\ and\ \citenamefont
  {Matsuo}}]{Kitsunezaki16}%
  \BibitemOpen
  \bibfield  {author} {\bibinfo {author} {\bibfnamefont {S.}~\bibnamefont
  {Kitsunezaki}}, \bibinfo {author} {\bibfnamefont {A.}~\bibnamefont
  {Nakahara}},\ and\ \bibinfo {author} {\bibfnamefont {Y.}~\bibnamefont
  {Matsuo}},\ }\href@noop {} {\bibfield  {journal} {\bibinfo  {journal}
  {Europhys. Lett.}\ }\textbf {\bibinfo {volume} {114}},\ \bibinfo {pages}
  {64002} (\bibinfo {year} {2016})}\BibitemShut {NoStop}%
\bibitem [{\citenamefont {Morita}\ and\ \citenamefont
  {Otsuki}(2021)}]{Morita21}%
  \BibitemOpen
  \bibfield  {author} {\bibinfo {author} {\bibfnamefont {J.}~\bibnamefont
  {Morita}}\ and\ \bibinfo {author} {\bibfnamefont {M.}~\bibnamefont
  {Otsuki}},\ }\href@noop {} {\bibfield  {journal} {\bibinfo  {journal} {Eur. Phys. J. E}\ }\textbf {\bibinfo {volume} {44}},\ \bibinfo
  {pages} {106} (\bibinfo {year} {2021})}\BibitemShut {NoStop}%
\bibitem [{Note1()}]{Note1}%
  \BibitemOpen
  \bibinfo {note} {The density of the agar was increased from
  the value $4\protect \tmspace +\thickmuskip {.2777em}\protect \mathrm {g/l}$
  used in Ref.~\cite {Kitsunezaki17} in order to realize solidification in samples with smaller
  solid volume fractions in the initial stages of desiccation.}\BibitemShut
  {Stop}%
\bibitem [{Note2()}]{Note2}%
  \BibitemOpen
  \bibinfo {note} {However, we found that if we stored paste
  containing both CsCl and agar for more than a few days before the
  experiments, the memory effect of shaking was lessened and in some cases
  disappeared. In this case, the samples tended to become fragile and exhibit
  irregular crack patterns. For this reason, we used only fresh paste stored
  for less than $12 \mathrm{h}$ hours in preparing samples}\BibitemShut {NoStop}%
\bibitem [{Ues()}]{Uesugi}%
  \BibitemOpen
  \href@noop {} {}\bibinfo {note} {Reconstruction software for $\mu$CT in
  SPring-8, available at http://www-bl20.spring8.or.jp/xct/index-e.html}\BibitemShut
  {NoStop}%
\bibitem [{\citenamefont {Schindelin}\ \emph {et~al.}(2012)\citenamefont
  {Schindelin}, \citenamefont {Arganda-Carreras}, \citenamefont {Frise},
  \citenamefont {Kaynig}, \citenamefont {Longair}, \citenamefont {Pietzsch},
  \citenamefont {Preibisch}, \citenamefont {Rueden}, \citenamefont {Saalfeld},
  \citenamefont {Schmid} \emph {et~al.}}]{Fiji}%
  \BibitemOpen
  \bibfield  {author} {\bibinfo {author} {\bibfnamefont {J.}~\bibnamefont
  {Schindelin}}, \bibinfo {author} {\bibfnamefont {I.}~\bibnamefont
  {Arganda-Carreras}}, \bibinfo {author} {\bibfnamefont {E.}~\bibnamefont
  {Frise}}, \bibinfo {author} {\bibfnamefont {V.}~\bibnamefont {Kaynig}},
  \bibinfo {author} {\bibfnamefont {M.}~\bibnamefont {Longair}}, \bibinfo
  {author} {\bibfnamefont {T.}~\bibnamefont {Pietzsch}}, \bibinfo {author}
  {\bibfnamefont {S.}~\bibnamefont {Preibisch}}, \bibinfo {author}
  {\bibfnamefont {C.}~\bibnamefont {Rueden}}, \bibinfo {author} {\bibfnamefont
  {S.}~\bibnamefont {Saalfeld}}, \bibinfo {author} {\bibfnamefont
  {B.}~\bibnamefont {Schmid}}, \emph {et~al.},\ }\href@noop {} {\bibfield
  {journal} {\bibinfo  {journal} {Nat. Methods}\ }\textbf {\bibinfo {volume}
  {9}},\ \bibinfo {pages} {676} (\bibinfo {year} {2012})}. \ \bibinfo {note}
  {FIJI, a public domain image processing program based on IMAGEJ
  (https://fiji.sc), was used for image analysis in this paper}\BibitemShut
  {NoStop}%
\bibitem [{\citenamefont {Doube}\ \emph {et~al.}(2010)\citenamefont {Doube},
  \citenamefont {K{\l}osowski}, \citenamefont {Arganda-Carreras}, \citenamefont
  {Cordeli{\`e}res}, \citenamefont {Dougherty}, \citenamefont {Jackson},
  \citenamefont {Schmid}, \citenamefont {Hutchinson},\ and\ \citenamefont
  {Shefelbine}}]{BoneJ}%
  \BibitemOpen
  \bibfield  {author} {\bibinfo {author} {\bibfnamefont {M.}~\bibnamefont
  {Doube}}, \bibinfo {author} {\bibfnamefont {M.~M.}\ \bibnamefont
  {K{\l}osowski}}, \bibinfo {author} {\bibfnamefont {I.}~\bibnamefont
  {Arganda-Carreras}}, \bibinfo {author} {\bibfnamefont {F.~P.}\ \bibnamefont
  {Cordeli{\`e}res}}, \bibinfo {author} {\bibfnamefont {R.~P.}\ \bibnamefont
  {Dougherty}}, \bibinfo {author} {\bibfnamefont {J.~S.}\ \bibnamefont
  {Jackson}}, \bibinfo {author} {\bibfnamefont {B.}~\bibnamefont {Schmid}},
  \bibinfo {author} {\bibfnamefont {J.~R.}\ \bibnamefont {Hutchinson}},\ and\
  \bibinfo {author} {\bibfnamefont {S.~J.}\ \bibnamefont {Shefelbine}},\
  }\href@noop {} {\bibfield  {journal} {\bibinfo  {journal} {Bone}\ }\textbf
  {\bibinfo {volume} {47}},\ \bibinfo {pages} {1076} (\bibinfo {year}
  {2010})}.\ \bibinfo {note} {We used this 3D particle analyzer plugin of IMAGEJ  (https://bonej.org/).}\BibitemShut {Stop}%
\bibitem [{\citenamefont {Ollion}\ \emph {et~al.}(2013)\citenamefont {Ollion},
  \citenamefont {Cochennec}, \citenamefont {Loll}, \citenamefont {Escud{\'e}},\
  and\ \citenamefont {Boudier}}]{3DDistanceMap}%
  \BibitemOpen
  \bibfield  {author} {\bibinfo {author} {\bibfnamefont {J.}~\bibnamefont
  {Ollion}}, \bibinfo {author} {\bibfnamefont {J.}~\bibnamefont {Cochennec}},
  \bibinfo {author} {\bibfnamefont {F.}~\bibnamefont {Loll}}, \bibinfo {author}
  {\bibfnamefont {C.}~\bibnamefont {Escud{\'e}}},\ and\ \bibinfo {author}
  {\bibfnamefont {T.}~\bibnamefont {Boudier}},\ }\href@noop {} {\bibfield
  {journal} {\bibinfo  {journal} {Bioinformatics}\ }\textbf {\bibinfo {volume}
  {29}},\ \bibinfo {pages} {1840} (\bibinfo {year} {2013})}.\ \bibinfo {note}
  {We used this 3D viewer plugin of IMAGEJ}\BibitemShut {NoStop}%
\bibitem [{\citenamefont {Weinberger}(1999)}]{Weinberger99}%
  \BibitemOpen
  \bibfield  {author} {\bibinfo {author} {\bibfnamefont {R.}~\bibnamefont
  {Weinberger}},\ }\href@noop {} {\bibfield  {journal} {\bibinfo  {journal} {J.
  Struct. Geol.}\ }\textbf {\bibinfo {volume} {21}},\ \bibinfo {pages} {379}
  (\bibinfo {year} {1999})}\BibitemShut {NoStop}%
\bibitem [{\citenamefont {Weinberger}(2001)}]{Weinberger01}%
  \BibitemOpen
  \bibfield  {author} {\bibinfo {author} {\bibfnamefont {R.}~\bibnamefont
  {Weinberger}},\ }\href@noop {} {\bibfield  {journal} {\bibinfo  {journal}
  {Geol. Soc. Amer. Bull.}\ }\textbf {\bibinfo {volume} {113}},\ \bibinfo
  {pages} {20} (\bibinfo {year} {2001})}\BibitemShut {NoStop}%
\bibitem [{\citenamefont {Kitsunezaki}(2009)}]{Kitsunezaki09}%
  \BibitemOpen
  \bibfield  {author} {\bibinfo {author} {\bibfnamefont {S.}~\bibnamefont
  {Kitsunezaki}},\ }\href@noop {} {\bibfield  {journal} {\bibinfo  {journal}
  {J. Phys. Soc. Jpn.}\ }\textbf {\bibinfo {volume} {78}},\ \bibinfo {pages}
  {064801} (\bibinfo {year} {2009})}\BibitemShut {NoStop}%
\bibitem [{\citenamefont {Denn}\ \emph {et~al.}(2018)\citenamefont {Denn},
  \citenamefont {Morris},\ and\ \citenamefont {Bonn}}]{Denn18}%
  \BibitemOpen
  \bibfield  {author} {\bibinfo {author} {\bibfnamefont {M.~M.}\ \bibnamefont
  {Denn}}, \bibinfo {author} {\bibfnamefont {J.~F.}\ \bibnamefont {Morris}},\
  and\ \bibinfo {author} {\bibfnamefont {D.}~\bibnamefont {Bonn}},\ }\href@noop
  {} {\bibfield  {journal} {\bibinfo  {journal} {Soft Matter}\ }\textbf
  {\bibinfo {volume} {14}},\ \bibinfo {pages} {170} (\bibinfo {year}
  {2018})}\BibitemShut {NoStop}%
\bibitem [{\citenamefont {Roch{\'e}}\ \emph {et~al.}(2011)\citenamefont
  {Roch{\'e}}, \citenamefont {Kellay},\ and\ \citenamefont {Stone}}]{Roche11}%
  \BibitemOpen
  \bibfield  {author} {\bibinfo {author} {\bibfnamefont {M.}~\bibnamefont
  {Roch{\'e}}}, \bibinfo {author} {\bibfnamefont {H.}~\bibnamefont {Kellay}},\
  and\ \bibinfo {author} {\bibfnamefont {H.~A.}\ \bibnamefont {Stone}},\
  }\href@noop {} {\bibfield  {journal} {\bibinfo  {journal} {Phys. Rev. Lett.}\ }\textbf {\bibinfo {volume} {107}},\ \bibinfo {pages} {134503}
  (\bibinfo {year} {2011})}\BibitemShut {NoStop}%
\bibitem [{\citenamefont {Guazzelli}\ and\ \citenamefont
  {Pouliquen}(2018)}]{Guazzelli18}%
  \BibitemOpen
  \bibfield  {author} {\bibinfo {author} {\bibfnamefont {{\'E}.}~\bibnamefont
  {Guazzelli}}\ and\ \bibinfo {author} {\bibfnamefont {O.}~\bibnamefont
  {Pouliquen}},\ }\href@noop {} {\bibfield  {journal} {\bibinfo  {journal}
  {Journal of Fluid Mechanics}\ }\textbf {\bibinfo {volume} {852}},\ P1 (\bibinfo
  {year} {2018})}\BibitemShut {NoStop}%
\bibitem [{\citenamefont {Varga}\ \emph {et~al.}(2019)\citenamefont {Varga},
  \citenamefont {Grenard}, \citenamefont {Pecorario}, \citenamefont {Taberlet},
  \citenamefont {Dolique}, \citenamefont {Manneville}, \citenamefont {Divoux},
  \citenamefont {McKinley},\ and\ \citenamefont {Swan}}]{Varga19}%
  \BibitemOpen
  \bibfield  {author} {\bibinfo {author} {\bibfnamefont {Z.}~\bibnamefont
  {Varga}}, \bibinfo {author} {\bibfnamefont {V.}~\bibnamefont {Grenard}},
  \bibinfo {author} {\bibfnamefont {S.}~\bibnamefont {Pecorario}}, \bibinfo
  {author} {\bibfnamefont {N.}~\bibnamefont {Taberlet}}, \bibinfo {author}
  {\bibfnamefont {V.}~\bibnamefont {Dolique}}, \bibinfo {author} {\bibfnamefont
  {S.}~\bibnamefont {Manneville}}, \bibinfo {author} {\bibfnamefont
  {T.}~\bibnamefont {Divoux}}, \bibinfo {author} {\bibfnamefont {G.~H.}\
  \bibnamefont {McKinley}},\ and\ \bibinfo {author} {\bibfnamefont {J.~W.}\
  \bibnamefont {Swan}},\ }\href@noop {} {\bibfield  {journal} {\bibinfo
  {journal} {Proc. Natl. Acad. Sci. U.S.A.}\ }\textbf
  {\bibinfo {volume} {116}},\ \bibinfo {pages} {12193} (\bibinfo {year}
  {2019})}\BibitemShut {NoStop}%
\bibitem [{\citenamefont {Cheng}\ \emph {et~al.}(2012)\citenamefont {Cheng},
  \citenamefont {Xu}, \citenamefont {Rice}, \citenamefont {Dinner},\ and\
  \citenamefont {Cohen}}]{Cheng12}%
  \BibitemOpen
  \bibfield  {author} {\bibinfo {author} {\bibfnamefont {X.}~\bibnamefont
  {Cheng}}, \bibinfo {author} {\bibfnamefont {X.}~\bibnamefont {Xu}}, \bibinfo
  {author} {\bibfnamefont {S.~A.}\ \bibnamefont {Rice}}, \bibinfo {author}
  {\bibfnamefont {A.~R.}\ \bibnamefont {Dinner}},\ and\ \bibinfo {author}
  {\bibfnamefont {I.}~\bibnamefont {Cohen}},\ }\href@noop {} {\bibfield
  {journal} {\bibinfo  {journal} {Proc. Natl. Acad. Sci. U.S.A.}\ }\textbf {\bibinfo {volume} {109}},\ \bibinfo {pages} {63}
  (\bibinfo {year} {2012})}\BibitemShut {NoStop}%
\bibitem [{\citenamefont {Olsson}\ and\ \citenamefont
  {Teitel}(2007)}]{Olsson07}%
  \BibitemOpen
  \bibfield  {author} {\bibinfo {author} {\bibfnamefont {P.}~\bibnamefont
  {Olsson}}\ and\ \bibinfo {author} {\bibfnamefont {S.}~\bibnamefont
  {Teitel}},\ }\href@noop {} {\bibfield  {journal} {\bibinfo  {journal}
  {Phys. Rev. Lett.}\ }\textbf {\bibinfo {volume} {99}},\ \bibinfo
  {pages} {178001} (\bibinfo {year} {2007})}\BibitemShut {NoStop}%
\bibitem [{\citenamefont {Li}\ \emph {et~al.}(2008)\citenamefont {Li},
  \citenamefont {Zhao},\ and\ \citenamefont {Liu}}]{Li08}%
  \BibitemOpen
  \bibfield  {author} {\bibinfo {author} {\bibfnamefont {S.}~\bibnamefont
  {Li}}, \bibinfo {author} {\bibfnamefont {L.}~\bibnamefont {Zhao}},\ and\
  \bibinfo {author} {\bibfnamefont {Y.}~\bibnamefont {Liu}},\ }\href@noop {}
  {\bibfield  {journal} {\bibinfo  {journal} {Comput. Mater. Contin.}\ }\textbf {\bibinfo {volume} {7}},\ \bibinfo {pages} {109}
  (\bibinfo {year} {2008})}\BibitemShut {NoStop}%
\bibitem [{\citenamefont {Bormashenko}\ \emph
  {et~al.}(2009{\natexlab{a}})\citenamefont {Bormashenko}, \citenamefont
  {Stein}, \citenamefont {Pogreb},\ and\ \citenamefont
  {Aurbach}}]{Bormashenko09}%
  \BibitemOpen
  \bibfield  {author} {\bibinfo {author} {\bibfnamefont {E.}~\bibnamefont
  {Bormashenko}}, \bibinfo {author} {\bibfnamefont {T.}~\bibnamefont {Stein}},
  \bibinfo {author} {\bibfnamefont {R.}~\bibnamefont {Pogreb}},\ and\ \bibinfo
  {author} {\bibfnamefont {D.}~\bibnamefont {Aurbach}},\ }\href@noop {}
  {\bibfield  {journal} {\bibinfo  {journal} {J. Phys. Chem. C}\ }\textbf {\bibinfo {volume} {113}},\ \bibinfo {pages} {5568} (\bibinfo
  {year} {2009}{\natexlab{a}})}\BibitemShut {NoStop}%
\bibitem [{\citenamefont {Bormashenko}\ \emph
  {et~al.}(2009{\natexlab{b}})\citenamefont {Bormashenko}, \citenamefont
  {Pogreb}, \citenamefont {Whyman}, \citenamefont {Musin}, \citenamefont
  {Bormashenko},\ and\ \citenamefont {Barkay}}]{Bormashenko09b}%
  \BibitemOpen
  \bibfield  {author} {\bibinfo {author} {\bibfnamefont {E.}~\bibnamefont
  {Bormashenko}}, \bibinfo {author} {\bibfnamefont {R.}~\bibnamefont {Pogreb}},
  \bibinfo {author} {\bibfnamefont {G.}~\bibnamefont {Whyman}}, \bibinfo
  {author} {\bibfnamefont {A.}~\bibnamefont {Musin}}, \bibinfo {author}
  {\bibfnamefont {Y.}~\bibnamefont {Bormashenko}},\ and\ \bibinfo {author}
  {\bibfnamefont {Z.}~\bibnamefont {Barkay}},\ }\href@noop {} {\bibfield
  {journal} {\bibinfo  {journal} {Langmuir}\ }\textbf {\bibinfo {volume}
  {25}},\ \bibinfo {pages} {1893} (\bibinfo {year}
  {2009}{\natexlab{b}})}\BibitemShut {NoStop}%
\bibitem [{\citenamefont {Seto}\ \emph {et~al.}(2013)\citenamefont {Seto},
  \citenamefont {Mari}, \citenamefont {Morris},\ and\ \citenamefont
  {Denn}}]{Seto13}%
  \BibitemOpen
  \bibfield  {author} {\bibinfo {author} {\bibfnamefont {R.}~\bibnamefont
  {Seto}}, \bibinfo {author} {\bibfnamefont {R.}~\bibnamefont {Mari}}, \bibinfo
  {author} {\bibfnamefont {J.~F.}\ \bibnamefont {Morris}},\ and\ \bibinfo
  {author} {\bibfnamefont {M.~M.}\ \bibnamefont {Denn}},\ }\href@noop {}
  {\bibfield  {journal} {\bibinfo  {journal} {Phys. Rev. Lett.}\
  }\textbf {\bibinfo {volume} {111}},\ \bibinfo {pages} {218301} (\bibinfo
  {year} {2013})}\BibitemShut {NoStop}%
\bibitem [{\citenamefont {Vinutha}\ and\ \citenamefont
  {Sastry}(2016)}]{Vinutha16}%
  \BibitemOpen
  \bibfield  {author} {\bibinfo {author} {\bibfnamefont {H.}~\bibnamefont
  {Vinutha}}\ and\ \bibinfo {author} {\bibfnamefont {S.}~\bibnamefont
  {Sastry}},\ }\href@noop {} {\bibfield  {journal} {\bibinfo  {journal} {Nat. Phys.}\ }\textbf {\bibinfo {volume} {12}},\ \bibinfo {pages} {578}
  (\bibinfo {year} {2016})}\BibitemShut {NoStop}%
\bibitem [{\citenamefont {Koeze}\ \emph {et~al.}(2020)\citenamefont {Koeze},
  \citenamefont {Hong}, \citenamefont {Kumar},\ and\ \citenamefont
  {Tighe}}]{Koeze20}%
  \BibitemOpen
  \bibfield  {author} {\bibinfo {author} {\bibfnamefont {D.~J.}\ \bibnamefont
  {Koeze}}, \bibinfo {author} {\bibfnamefont {L.}~\bibnamefont {Hong}},
  \bibinfo {author} {\bibfnamefont {A.}~\bibnamefont {Kumar}},\ and\ \bibinfo
  {author} {\bibfnamefont {B.~P.}\ \bibnamefont {Tighe}},\ }\href@noop {}
  {\bibfield  {journal} {\bibinfo  {journal} {Phys. Rev. Research}\
  }\textbf {\bibinfo {volume} {2}},\ \bibinfo {pages} {032047(R)} (\bibinfo {year}
  {2020})}\BibitemShut {NoStop}%
\end{thebibliography}



%

\end{document}